\documentclass[a4paper,11pt]{article}
\pdfoutput=1 

\usepackage{jheppub} 

\usepackage{graphicx}

\usepackage{MnSymbol}
\usepackage{empheq}
\usepackage{slashed} 

\usepackage{hyperref}

\usepackage{tikz}

\usetikzlibrary{decorations}

\title{\boldmath Poincar\'e Constraints on Celestial Amplitudes}

\author{Y. T. Albert Law,$^1$}
\author{Michael Zlotnikov,$^1$}

\affiliation{$^1$ Department of Physics, Center for Theoretical Physics,
Columbia University, 538 West 120th Street, New York, NY 10027, USA.}
\emailAdd{yal2109@columbia.edu}
\emailAdd{mz2737@columbia.edu}

\abstract{\\
The functional structure of celestial amplitudes as constrained by Poincar\'e symmetry is investigated in $2,3,$ and $4$-point cases for massless external particles of various spin, as well as massive external scalars. Functional constraints and recurrence relations are found (akin to the findings in \href[pdfnewwindow=true]{https://arxiv.org/abs/1901.01622}{1901.01622}) that must be obeyed by the respective permissible correlator structures and function coefficients. In specific three-point cases involving massive scalars the resulting recurrence relations can be solved, e.g., reproducing purely from symmetry a three-point function coefficient known in the literature. Additionally, as a byproduct of the analysis, the three-point function coefficient for gluons in Minkowski signature is obtained from an amplitude map to the celestial sphere.
}

\begin{document} 
\maketitle
\flushbottom

\section{Introduction}
Foliating Minkowski space by hyperbolic slices to establish flat space holography was proposed some time ago by de Boer and Solodukhin \cite{deBoer:2003vf}.
The idea of flat space holography was picked up, e.g., in relating 4D soft theorems to conserved currents on the celestial sphere \cite{He:2015zea,Cheung:2016iub}.
Recent interest in flat space holography has been enhanced by the concrete proposal by Pasterski, Shao and Strominger (PSS) in \cite{Pasterski:2016qvg} to map the flat space plane wave basis to a conformal primary wave function basis on the celestial sphere via an integral transform, allowing to express four-dimensional scattering amplitudes in terms of two-dimensional correlator like objects (celestial amplitudes). The singularity structure of higher-point amplitudes was considered on the celestial sphere in \cite{Cardona:2017keg}. The basis of conformal primary wave functions for massless particles of spins zero, one, and two was derived in \cite{Pasterski:2017kqt}, establishing the map to the celestial sphere in these cases, after gauge fixing, to be given by Mellin transform. Following the PSS prescription, explicit examples of amplitudes were mapped to the celestial sphere for scalar scattering \cite{Pasterski:2016qvg,Lam:2017ofc,Banerjee:2017jeg,Nandan:2019jas}, gluon scattering \cite{Pasterski:2017ylz,Schreiber:2017jsr}, and stringy/graviton scattering \cite{Stieberger:2018edy,Puhm:2019zbl,Guevara:2019ypd}.
Modification of the PSS prescription, which makes the action of space-time translation simpler, has been proposed and investigated in \cite{Banerjee:2018gce,Banerjee:2018fgd,Banerjee:2019prz}.
Conformally soft behavior of operators on the celestial sphere was considered in \cite{Donnay:2018neh,Banerjee:2019aoy,Fan:2019emx,Pate:2019mfs,Nandan:2019jas,Adamo:2019ipt,Puhm:2019zbl,Guevara:2019ypd,Himwich:2019dug}. 
Conformal partial wave decomposition of some four-point amplitudes on the celestial sphere was discussed in \cite{Lam:2017ofc,Nandan:2019jas}.
The representation of Poincar\'e symmetry generators for massless particles on the celestial sphere was presented in \cite{Stieberger:2018onx}. BMS symmetry in the language of flat space holography was recently considered in \cite{Ball:2019atb}. And the OPE of the energy-momentum tensor (see \cite{Kapec:2016jld,Cheung:2016iub,Kapec:2017gsg}) with gauge boson operators on the celestial sphere was obtained in \cite{Fotopoulos:2019tpe}, demonstrating that gauge boson operators in fact transform as Virasoro primaries.

Since the integral transform of amplitudes to the celestial sphere is in general hard to perform, only a limited number of examples has been calculated so far. In this work we take a parallel approach and instead consider constraints imposed on generic celestial amplitudes by the underlying Poincar\'e symmetry. Our main motivations are two-fold:
\begin{itemize}
	\item First, we are looking to constrain the general non-perturbative structure of celestial amplitudes in order to aid future calculations of amplitudes on the celestial sphere.
	\item Second, we are interested in determining which features of the currently known celestial amplitude examples stem from the particular types of particles and theories under consideration, and which features are more generally to be attributed to the overarching symmetry.
\end{itemize}
Poincar\'e symmetry on the celestial sphere is realized non-trivially and imposes functional constraints on correlator structures. We employ the condition that celestial amplitudes must be annihilated by all Poincar\'e symmetry generators (conformal Ward identities and momentum conservation). Parametrization of Poincar\'e symmetry generators for massless particles is provided in \cite{Stieberger:2018onx}, while we additionally derive momentum generator parametrization for massive scalars, which allows us to investigate amplitudes with massless external particles of various spin, as well as massive external scalars.\footnote{Note that constraints found in \cite{Banerjee:2018gce}, while not without similarities, refer to a different functional basis.} 

In the following paragraphs we summarize the main results of this work.

In case when all external particles are massless, we find that two-, three- and four-point structures are required by Poincar\'e symmetry to be distribution-valued. In fact, we show that the two- and three-point structures are then constrained to vanish. 

Despite its vanishing on the support of the delta functions, in the three-point case we find that the general structure can be written\footnote{Where $w_{ij}=w_i-w_j$, $\bar w_{ij}=\bar w_i-\bar w_j$, and $h_i=\frac{\Delta_i+J_i}{2}$, $\bar h_i=\frac{\Delta_i-J_i}{2}$ with conformal dimension in continuous series representation $\Delta_i=1+i\lambda_i$ with $\lambda\in\mathbb{R}$ and spin $J_i$.}
\begin{align}
\begin{split}
A_3=C_{\bar h_1,\bar h_2,\bar h_3}^{h_1,h_2,h_3}&w_{12}^{1-h_1-h_2+h_3}w_{23}^{h_1-h_2-h_3}w_{31}^{1-h_1+h_2-h_3}
 \bar w_{12}^{1-\bar h_1-\bar h_2+\bar h_3}\bar w_{23}^{\bar h_1-\bar h_2-\bar h_3}\bar w_{31}^{1-\bar h_1+\bar h_2-\bar h_3}\\
 & \delta\left(w_{12}\right)\delta\left(\bar w_{12}\right)\delta\left(w_{13}\right)\delta\left(\bar w_{13}\right)\,,
\end{split}
\end{align}
while Poincar\'e symmetry implies that a non-trivial three-point function coefficient $C_{\bar h_1,\bar h_2,\bar h_3}^{h_1,h_2,h_3}$ must exist, which must satisfy a recurrence relation under shifts in conformal dimensions. For instance, the gluon OPE coefficient was previously found to involve the Euler beta function \cite{Fan:2019emx,Fotopoulos:2019tpe}. As a byproduct of our investigation, in appendix \ref{sec:gluon3pt} we calculate the gluon three-point function coefficient from an amplitude map to the celestial sphere and confirm that the beta function recurrence identity in this context is a consequence of the global translation invariance constraint, as mentioned in \cite{Strominger:NEStringMeeting2019}. However, note that due to the distribution-valuedness and vanishing of the three-point structure, the three-point function coefficient does not have a simple relation to the OPE coefficient in this case.

Perhaps the most interesting constraint we find, is the fact that a massless four-point amplitude on the celestial sphere can always be written as\footnote{Here $z$ and $\bar z$ are conformal cross ratios $z=\frac{w_{12}w_{34}}{w_{13}w_{24}}$ and $\bar z=\frac{\bar w_{12}\bar w_{34}}{\bar w_{13}\bar w_{24}}$.}
\begin{align}
\begin{split}
A_4=&\frac{\left(\frac{w_1-w_4}{w_1-w_3}\right)^{h_3-h_4}
   \left(\frac{w_2-w_4}{w_1-w_4}\right)^{h_1-h_2}}{\left(w_1-w_2\right)^{h_1+h_2} \left(w_3-w_4\right)^{h_3+h_4}} 
   \frac{\left(\frac{\bar{w}_1-\bar{w}_4}{\bar{w}_1-\bar{w}_3}\right)^{\bar{h}_3-\bar{h}_4}
   \left(\frac{\bar{w}_2-\bar{w}_4}{\bar{w}_1-\bar{w}_4}\right)^{\bar{h}_1-\bar{h}_2}}{\left(\bar{w}_1-\bar{w}_2\right)^{\bar{h}_1+\bar{h}_2}\left(\bar{w}_3-\bar{w}_4\right)^{\bar{h}_3+\bar{h}_4}}\\
& \delta\left(i\bar z - i z\right)(z-1)^{\frac{h_1-h_2-h_3+h_4}{2}}(\bar z-1)^{\frac{\bar h_1-\bar h_2-\bar h_3+\bar h_4}{2}}\tilde f_{\Delta_1,\Delta_2,\Delta_3,\Delta_4}^{J_1,J_2,J_3,J_4}(z ,\bar z),
\end{split}
\end{align}
with a specific class of functions $\tilde f$ that may depend either on $\sum_i\Delta_i$, or individual $\Delta_i$ contributions that are periodic. We show that reasonably behaved holomorphic periodic functions are not bounded on any complex vertical line in the complex plane of $\Delta_i$, and therefore may spoil the convergence of the inverse integral transform of celestial amplitudes back to Minkowski space \cite{Pasterski:2017kqt}.

In case when massive external scalars are present, the Lorentz subgroup constrains the two-, three- and four-point structures to have the same familiar form as usual CFT correlators. We find that the momentum generators then impose further constraints on the coefficients depending on conformal weights, as well as on the function of cross-ratio in the four-point case.

The two-point structure of two massive scalars is only non-vanishing when both masses are equal, and we explicitly solve for the corresponding two-point function coefficient. The two-point structure of a massless particle with a massive scalar is ruled out. 

For the three-point structure of two massless particles with a massive scalar, we find that it only exists if the two massless particles have the same spin. We discover that the recurrence relations imposed on the three-point function coefficient by Poincar\'e symmetry are solved by a specific class of functions, which, e.g., in its simplest form reproduces the three-point function coefficient of two massless scalars with a massive scalar \cite{Lam:2017ofc} purely from symmetry considerations.

We find that the three-point function coefficient for a massless scalar with two massive scalars of different mass must obey three non-trivial three-term recurrence relations. In the case when the massless particle is spinning, we find a different set of recurrence relations, forcing the two massive scalars to have the same mass. Furthermore, the three-point function coefficient becomes anti-periodic in conformal dimensions, which makes it unbounded on any complex vertical line, such that the convergence of the inverse integral transform of the celestial amplitude to Minkowski space may not be well defined. This is in line with the fact that the equality of the two masses enforces a pathological kinematic configuration in Minkowski space, so that the amplitude is discarded.

We uncover that the three-point function coefficient for three massive scalars of different mass must obey three non-trivial four-term recurrence relations.

In case of four-point structures with at least one massive scalar, we find that the function of cross-ratios must satisfy second order differential equations, additionally subject to recurrence shifts in the conformal weights of external particles.

This work is organized as follows. In section \ref{sec:formalism} we recall the formalism for mapping amplitudes of massless and massive particles to the celestial sphere. In section \ref{sec:poincare} we list Poincar\'e symmetry generators and algebra acting on the celestial sphere. Section \ref{sec:massless} details the derivation of Poincar\'e symmetry constraints on two-, three- and four-point massless correlator structures, while section \ref{sec:massive} repeats that exercise for correlator structures also involving massive scalars. We offer some discussion in section \ref{sec:discussion}. Finally, appendix \ref{sec:gluon3pt} describes the calculation of the gluon three-point celestial amplitude starting with Minkowski signature in the bulk, while appendix \ref{A4shifts} describes the derivation of the class of functions solving the massless four-point structure recurrence relations.

\section{Formalism and conventions}
\label{sec:formalism}
We recall the formalism of mapping a massive scalar, or massless scalar, gluon or graviton plane wave solution to the celestial sphere.

In the case of the massive scalar, the map involves an integral transform over a hyperbolic slice $H_3$ of Minkowski space corresponding to the constant mass squared $p_j^2=-m_j^2$ hyper surface for external particle momenta \cite{Pasterski:2016qvg}. 
An $H_3$ slice in Poincar\'e coordinates $y,z,\bar z$ has the metric
\begin{align}
ds_{H_3}^2=\frac{dy^2+dz d\bar z}{y^2}~~~\text{with}~~~0<y<\infty\,,\text{ and}~z=(\bar z)^*\in\mathbb{C}\,,
\end{align}
which has the $SL(2,\mathbb{C})$ isometry
\begin{align}
z\to \frac{(a z+b)(\bar c\bar z+\bar d)+a\bar c y^2}{(c z+d)(\bar c \bar z+\bar d)+c \bar c y^2}~,~\bar z\to \frac{(\bar a \bar z+\bar b)(c  z+  d)+\bar a  c y^2}{(\bar c \bar z+\bar d)( c z+d)+c \bar c y^2}~,~y\to \frac{y}{(c z+d)(\bar c \bar z+\bar d)+c \bar c y^2},
\end{align}
where $a,b,c,d\in\mathbb{C}$ and $ad-bc=\bar a\bar d-\bar b\bar c=1$. The transformation parameters $a,b,c,d$ are the same that enter the corresponding $SL(2,\mathbb{C})$ M\"obius transformations 
\begin{align}
\label{Moebius}
w_i\to\frac{a w_i+b}{c w_i+d}~,~~\bar w_i\to\frac{\bar a \bar w_i+\bar b}{\bar c \bar w_i+\bar d}\,,
\end{align}
acting on complex coordinates $w_i,\bar w_i$ on the celestial sphere.

We also recall the embedding map for the (unit) momentum of a particle living on the hyperboloid $\hat p^\mu:H_3\to\mathbb{R}^{1,3}$ to be
\begin{align}
\hat p(y,z,\bar z)^\mu=\left(\frac{1+y^2+z\bar z}{2y},\frac{\bar z+z}{2y},i\frac{\bar z-z}{2y},\frac{1-y^2-z\bar z}{2y}\right).
\end{align}
In \cite{Pasterski:2016qvg} the plane wave basis of a mass $m$ scalar (outgoing or incoming $\pm$) was mapped to the conformal primary wave function basis with conformal dimension $\Delta$ on the celestial sphere (in the continuous series representation $\Delta=1+i\lambda$ with $\lambda\in\mathbb{R}$) via the integral transform
\begin{align}
\label{s0Phi}
\phi_{\pm,\Delta,m}(X^\mu;w,\bar w)&=\int_0^\infty\frac{dy}{y^3}\int dz d\bar z \,\left(\frac{y}{y^2+|w-z|^2}\right)^\Delta e^{\pm i m\hat p^\nu X_\nu},
\end{align}
where the terms to the power $\Delta$ are the scalar bulk to boundary propagator on $H_3$. The bulk to boundary propagator can be written as $(-q\cdot \hat p)^{-\Delta}$, while $q_j^\mu$ is a null direction
\begin{align}
q_j^\mu&=(1+w_j \bar w_j,~\bar w_j + w_j,~i(\bar w_j-w_j),~1-w_j \bar w_j),
\end{align}
pointing to the celestial sphere.

The map of a Minkowski space amplitude of massive scalars $\mathcal{A}_n$ to the celestial sphere is then given by
\begin{align}
\label{Amass}
A_n=\left(\prod_i^n \int \frac{dy_i}{y_i^3}dz_i d\bar z_i \, (-q_i\cdot \hat p_i)^{-\Delta_i}\right)\mathcal{A}_n.
\end{align}
Similarly, in \cite{Pasterski:2017kqt} making use of a specific gauge, it was shown that amplitudes $\mathcal{A}_n$ of massless particles of spin zero, one, and two are mapped to the celestial sphere via the Mellin transform
\begin{align}
\label{Anomass}
A_n=\left(\prod_i^n \int_0^\infty d\omega_i \, \omega_i^{\Delta_i-1}\right)\mathcal{A}_n,
\end{align}
where incoming or outgoing massless particle momenta are $p_j^\mu=\pm\omega_j q_j^\mu$. Prescriptions (\ref{Amass}) and (\ref{Anomass}) are to be mixed appropriately, depending on the spin and mass of the respective external particles in amplitude $\mathcal{A}_n$.

\section{Poincar\'e generators and algebra}
\label{sec:poincare}
\paragraph{Poincar\'e algebra:}$~$\\
The Poicar\'e algebra is defined by the commutation relations
\begin{align}
\label{MMalg}
[M^{\mu\nu},M^{\rho\sigma}]=-i\left(\eta^{\mu\sigma}M^{\nu\rho}+\eta^{\nu\rho}M^{\mu\sigma}-\eta^{\mu\rho}M^{\nu\sigma}-\eta^{\nu\sigma}M^{\mu\rho}\right),
\end{align}
where $\eta^{\mu\nu}=\text{diag}(-1,1,1,1)$, and 
\begin{align}
\label{MPalg}
[M^{\mu\nu},P^{\rho}]=i\left(\eta^{\mu\rho}P^{\nu}-\eta^{\nu\rho}P^{\mu}\right)~~~~~~,~~~~~~[P^{\mu},P^{\nu}]=0\,,
\end{align}
where $M^{\mu\nu}$ are Lorentz generators and $P^\mu$ are translation (momentum) generators. 

\paragraph{Lorentz generators:}$~$\\
Lorentz generators $M^{\mu\nu}=-M^{\nu\mu}$ for massless or massive particles are given by \cite{Stieberger:2018onx}
\begin{align}
\label{lorentzgen}
\begin{split}
M^{01}&=\frac{i}{2}  \left(\left(\bar{w}^2-1\right)
   \bar\partial+\left(w^2-1\right)
   \partial+2 \left(\bar{h} \bar{w}+h w\right)
   \right)\,,\\
	M^{02}&=-\frac{1}{2} \left(\left(\bar{w}^2+1\right)
   \bar\partial-\left(w^2+1\right)
   \partial+2\left( \bar{h} \bar{w}- h w\right)
   \right)\,,\\
	M^{03}&=i \left(\bar{w} \bar \partial+w
   \partial+\bar{h}+h
   \right) \,,\\
	M^{12}&=-\bar{w} \bar \partial+w
   \partial-\bar{h}+h
   \,,\\
	M^{13}&=\frac{i}{2}  \left(\left(\bar{w}^2+1\right)
   \bar\partial+\left(w^2+1\right)
   \partial+2 \left(\bar{h} \bar{w}+h w\right)
   \right)\,,\\
	M^{23}&=-\frac{1}{2} \left(\left(\bar{w}^2-1\right)
   \bar\partial-\left(w^2-1\right)
   \partial+2\left( \bar{h} \bar{w}- h w\right)
   \right)\,,
	\end{split}
\end{align}
where $\partial=\frac{\partial}{\partial w}$ and $\bar \partial=\frac{\partial}{\partial \bar w}$. 

\paragraph{Massless momentum generator:}$~$\\
Translation generators (momenta) for massless particles act on massless states on the celestial sphere as follows \cite{Stieberger:2018onx}:
\begin{align}
\label{m0P}
P^\mu=q^\mu e^{\big(\partial_{h}+\partial_{{\bar h}}\big)/2},
\end{align}
where $\partial_x=\frac{\partial}{\partial x}$, and $h=\frac{\Delta+J}{2},\,\bar h=\frac{\Delta-J}{2}$ are the holomorphic and anti-holomorphic conformal weights of the respective particle on the celestial sphere (with spin $J$).

\paragraph{Massive scalar momentum generator:}$~$\\
Analogously to the massless case (\ref{m0P}), we provide a representation of massive momentum generators dependent only on celestial sphere coordinates $w,\bar w$, and conformal dimensions. As can be directly verified, the following operator representation indeed has the correct momentum component eigenvalues $m\hat p^\mu$ when acting on $(-q\cdot \hat p)^{-\Delta}$ within the massive scalar conformal primary wave function (\ref{s0Phi}):
\begin{align}
\label{PmS}
P^\mu=&\frac{m}{2}\left(\left(\partial_{w}\partial_{\bar w}q^\mu+\frac{(\partial_{\bar w} q^\mu)\partial_{ w}+(\partial_{w} q^\mu)\partial_{\bar w}}{\Delta-1}+\frac{ q^\mu\partial_{ w}\partial_{\bar w}}{(\Delta-1)^2}\right)e^{-\partial_{\Delta} }+\frac{\Delta q^\mu }{\Delta-1}e^{\partial_{\Delta} }\right),
\end{align}
with $m$ being the mass of the respective particle.

It is also straightforward to check explicitly that the operators (\ref{m0P}) or (\ref{PmS}) together with (\ref{lorentzgen}) correctly close the Poincar\'e algebra (\ref{MMalg}),(\ref{MPalg}). However, in the massive case (\ref{PmS}) this is only true if the conformal weights have no spin $h=\bar h=\frac{\Delta}{2}$. This demonstrates that the massive momentum generator representation on the celestial sphere is spin dependent.

\section{Poincar\'e constraints on massless 2,3 and 4-point structures}
\label{sec:massless}
Poincar\'e symmetry implies that all algebra generators have to annihilate the $n$-point amplitude structures on the celestial sphere by conformal Ward identity and momentum conservation constraints:
\begin{align}
\sum_{j=1}^n \epsilon_j P_j^{\mu}A(1,2,...,n)=0~~~~~~,~~~~~~\sum_{j=1}^n M_j^{\mu\nu}A(1,2,...,n)=0,
\end{align}
where $\epsilon_j=\pm 1$ depending on whether particle $j$ is outgoing or incoming.

 In the following we consider the cases $n=2,3,4$, for which the symmetry produces special constraints. We take an agnostic approach, seeking to discover constraints resulting purely from the above symmetry relations only, without imposing additional external knowledge by hand.

It turns out that the action of the massless momentum generators forces the $n=2,3,4$ celestial amplitudes to be distribution-valued. To see that, note that the equation $\sum_{j}\epsilon_j P_j^{\mu}A=0$ can be understood as a linear set of equations for $n$ unknowns $X_j\equiv \epsilon_j\exp(\frac{1}{2}\partial_{h_j}+\frac{1}{2}\partial_{\bar h_j})A$, which in matrix notation can be written as
\begin{align}
Q\cdot X=0~~~\text{with}~~~Q=(q_1^\mu q_2^\mu...q_n^\mu).
\end{align}
As is well known, such a linear set of equations can have non-trivial solutions for $n\leq 4$ only if the $4\times n$ matrix $Q$ has appropriately reduced rank. As a consequence, determinants of all maximal minors of matrix $Q$ must vanish. Such constraints reduce the regions of values the coordinates $w_j,\bar w_j$ can take on, which will be parametrized by Dirac delta distributions.

\subsection{Two-point structure}
\paragraph{Maximal minor determinant constraints}$~$\\
It is easy to check that the determinants of all six $2\times 2$ minors of matrix $Q=(q_1^\mu q_2^\mu)$ vanish only if the two points on the celestial sphere coincide
\begin{align}
w_1=w_2~~~,~~~\bar w_1 = \bar w_2.
\end{align}
This is in line with momentum conservation expectation in Minkowski space, stating that ingoing and outgoing states in a two point process should be collinear. Therefore, any valid two-point structure $A_2$ on the celestial sphere must be proportional to a product of delta functions imposing the above constraints:\footnote{Since the delta function arguments are complex conjugates of each other, they are to be understood as setting the real and the imaginary part to zero, which may be more transparent in the alternative linear combination of arguments: $\delta\left(w_1- w_2\right)\delta\left(\bar w_1- \bar w_2\right)\leftrightarrow\delta\left(\frac{w_1+\bar w_1}{2}-\frac{w_2+\bar w_2}{2}\right)\delta\left(\frac{i(\bar w_1-w_1)}{2}-\frac{i(\bar w_2-w_2)}{2}\right)$. }
\begin{align}
A_2&=g_{\bar h_1,\bar h_2}^{h_1,h_2}(w_1,\bar w_1,w_2,\bar w_2)\cdot \delta\left(w_1- w_2\right)\delta\left(\bar w_1- \bar w_2\right),
\end{align}
where $g$ is so far a generic function of its arguments.

\paragraph{Lorentz invariance constraints}$~$\\
Next we impose the Lorentz symmetry constraints. As one part of Lorentz invariance, the function $g$ should be invariant under global 2D translations on the celestial sphere. Therefore, we conclude that $g$ actually depends on differences $w_1-w_2$ and $\bar w_1-\bar w_2$ only, so that we take the general ansatz
\begin{align}
A_2=C^{h_1,h_2}_{\bar h_1,\bar h_2}(w_1-w_2)^p(\bar w_1-\bar w_2)^{\bar p} \delta\left(w_1- w_2\right)\delta\left(\bar w_1- \bar w_2\right),
\end{align}
where $p,\bar p$ depend on $h_1,h_2$ and $\bar h_1,\bar h_2$, and $C^{h_1,h_2}_{\bar h_1,\bar h_2}$ is a two-point function coefficient.

On the celestial sphere, the annihilation of a correlation function of primary fields by symmetry generators in the context of conformal Ward identities is equivalent to the condition that the correlator must have the correct primary-field transformation weight under M\"obius transformation (\ref{Moebius}).
The two-point structure $A_2$ must transform as
\begin{align}
A_2 \to (c w_1+d)^{2h_1}(c w_2+d)^{2h_2} (\bar c \bar w_1+\bar d)^{2\bar h_1}(\bar c \bar w_2+\bar d)^{2\bar h_2} A_2 .
\end{align}
It is straightforward to see that this transformation weight can be realized by our ansatz $A_2$ only if we demand
\begin{align}
h_1=h_2=h,~\bar h_1=\bar h_2=\bar h\,,~~~\text{and}~~~ p=1-2h\,,~\bar p=1-2\bar h\,,
\end{align}
so that analogously to the usual CFT case \cite{DiFrancesco:1997nk} the $w_i,\bar w_i$ dependence is completely fixed by symmetry, and $A_2$ becomes
\begin{align}
\label{A200}
A_2=C^{h_1,h_2}_{\bar h_1,\bar h_2}\frac{\hat\delta(h_1-h_2)\hat\delta(\bar h_1-\bar h_2)}{(w_1-w_2)^{h_1+h_2-1}(\bar w_1-\bar w_2)^{\bar h_1+\bar h_2-1}}\delta\left(w_1- w_2\right)\delta\left(\bar w_1- \bar w_2\right),
\end{align}
 where we define a finite-valued delta function
\begin{align}
\label{findelta}
\hat\delta(x)\equiv\lim_{\epsilon\to 0}\int_{-\epsilon}^\epsilon dy \delta(y+x)=\begin{cases}
 1 & x=0 \\
 0 & x\neq 0
\end{cases},
\end{align}
to ensure the conformal weight equality $h_1=h_2$, $\bar h_1=\bar h_2$.\footnote{We must keep $h_1,\bar h_1$ and $h_2,\bar h_2$ formally apart, since global momentum generators act on these conformal weights separately. This is a clear difference compared to usual CFT.} From current considerations it is unclear whether the two-point structure may diverge when the conformal weights are equal; in such a case we can use a combination of Dirac and Kronecker deltas instead
\begin{align}
\hat\delta(h_1-h_2)\hat\delta(\bar h_1-\bar h_2)~~\to~~\delta(\Delta_1-\Delta_2)\delta_{J_1,J_2}.
\end{align}
This does not change the conclusions from further constraints on the two-point function coefficient.

As a cross-check we verify that Lorentz generators annihilate the two-point structure as well: $\sum_i M_i^{\mu\nu}A_2=0$. Since we have established that $A_2$ is a distribution, it is clear that derivatives hitting the delta functions will have to be evaluated in a distributional sense $f(x)\partial_x \delta(x)=-\delta(x)\partial_x f(x)$. To arrive at a correct result after such partial integration, it is important to choose a proper normalization for the operator $\sum_i M_i^{\mu\nu}$. To that end, we \textit{define} the action of the operator on $A_2$ to be
\begin{align}
\frac{1}{g_{\bar h_1,\bar h_2}^{h_1,h_2}(w_1,\bar w_1,w_2,\bar w_2)}\sum_i M_i^{\mu\nu}A_2=0,
\end{align}
where we divide out the continuous dependence on $w_1,w_2$ that appears in $A_2$, so that the equation formally isolates only the eigenvalue of $\sum_i M_i^{\mu\nu}$ next to the delta functions.

Resolving delta function derivatives in a distributional sense as mentioned above:
\begin{align}
f(...)\delta'(w_1-w_2)=-\delta(w_1-w_2)\partial_{w_1-w_2}f(...)=-\delta(w_1-w_2)\frac{\partial_{w_1}-\partial_{w_2}}{2}f(...)
\end{align}
and similarly for $\delta'(\bar w_1-\bar w_2)$, we confirm that the two-point structure is properly annihilated.

\paragraph{Global translation invariance constraints}$~$\\
 The equations $\sum_{j} \epsilon_j P_j^{\mu}A_2=0$ (with $\epsilon_1=-\epsilon_2$) create two independent constraints for $A_2$ in (\ref{A200}), since each operator $e^{(\partial_{h_j}+\partial_{{\bar h}_j})/2}$ shifts the arguments of the modified delta functions such that terms proportional to $\hat\delta(h_{12}\pm\frac{1}{2})$ appear, which cannot be canceled against each other.\footnote{Here and in the following we sometimes abbreviate differences of quantities with indices as $x_{ij}\equiv x_i-x_j$.} The powers in $w_{12}$ and $\bar w_{12}$ do not develop a positive real part overall, so that $\delta(w_{12})\delta(\bar w_{12})$ does not reduce the result to zero, thus the only way to satisfy the annihilation constraint is to demand
\begin{align}
C^{h_1+\frac{1}{2},h_2}_{\bar h_1+\frac{1}{2},\bar h_2}=0~~~,~~~C^{h_1,h_2+\frac{1}{2}}_{\bar h_1,\bar h_2+\frac{1}{2}}=0.
\end{align}
When a shift of the argument of a function makes it vanish and the function domain is unbounded, it means that the function itself is zero. Therefore, we conclude that in general
\begin{align}
C^{h_1,h_2}_{\bar h_1,\bar h_2}=0,
\end{align}
such that a massless two-point structure is ruled out by Poincar\'e symmetry. 

\subsection{Three-point structure}
\paragraph{Maximal minor determinant constraints}$~$\\
Demanding that determinants of all four $3\times 3$ minors of matrix $Q=(q_1^\mu q_2^\mu q_3^\mu)$ vanish, and fully reducing these constraints leads to four branches of solutions. The first branch demands the coincidence of all three complex points, e.g.,
\begin{align}
w_1=w_2~~~,~~~w_1=w_3~~~(\text{which implicitly also implies}~~~\bar w_1 = \bar w_2~~~,~~~\bar w_1 = \bar w_3),
\end{align}
while three more branches demand the coincidence of only pairs of points
\begin{align}
w_i=w_j~~~,~~~\bar w_i = \bar w_j~~~~~\text{for}~i\neq j~\text{with}~ i,j\in\{1,2,3\}.
\end{align}
From momentum conservation we have the intuition that once two out of three momenta become collinear, the third must become collinear as well; so that we pick the first branch to proceed. Therefore, we conclude that a generic massless three-point structure $A_3$ on the celestial sphere must be proportional to a product of delta functions imposing the first branch constraints, e.g.,\footnote{Once again, to make sense of the complex valued arguments of the delta functions, we may alternatively consider the real valued linear combinations of arguments $\delta\left(\frac{w_1+\bar w_1}{2}-\frac{w_3+\bar w_3}{2}\right)\delta\left(\frac{i(\bar w_1-w_1)}{2}-\frac{i(\bar w_3-w_3)}{2}\right)\delta\left(\frac{w_2+\bar w_2}{2}-\frac{w_1+\bar w_1}{2}\right)\delta\left(\frac{i(\bar w_2-w_2)}{2}-\frac{i(\bar w_1-w_1)}{2}\right)$.}
\begin{align}
A_3&=g_{\bar h_1,\bar h_2,\bar h_3}^{h_1,h_2,h_3}(w_1,\bar w_1,w_2,\bar w_2,w_3,\bar w_3)\cdot \delta\left(w_1-w_2\right)\delta\left(\bar w_1-\bar w_2\right)\delta\left(w_1-w_3\right)\delta\left(\bar w_1-\bar w_3\right),
\end{align}
where $g$ for now is an arbitrary function of its arguments.

\paragraph{Lorentz invariance constraints}$~$\\
We proceed to impose the Lorentz symmetry constraints. As before, due to 2D translation invariance on the celestial sphere, the three-point structure may depend only on differences of coordinates $w_i-w_j$ and $\bar w_i-\bar w_j$. This leads to the general ansatz
\begin{align}
A_3=C_{\bar h_1,\bar h_2,\bar h_3}^{h_1,h_2,h_3}&(w_1-w_2)^{p_1}(w_2-w_3)^{p_2}(w_3-w_1)^{p_3}
 (\bar w_1-\bar w_2)^{\bar p_1}(\bar w_2-\bar w_3)^{\bar p_2}(\bar w_3-\bar w_1)^{\bar p_3}\notag\\
 & \delta\left(w_1-w_2\right)\delta\left(\bar w_1-\bar w_2\right)\delta\left(w_1-w_3\right)\delta\left(\bar w_1-\bar w_3\right),
\end{align}
with a three-point function coefficient $C$. Demanding the three-point structure to have the correct primary-field transformation weight under M\"obius transformation
\begin{align}
A_3 \to \prod_i(c w_i+d)^{2h_i}(\bar c \bar w_i+\bar d)^{2\bar h_i} A_3,
\end{align}
fixes the powers $p_i,\bar p_i$ uniquely analogously to the usual CFT case \cite{DiFrancesco:1997nk}, so that we obtain the final result
\begin{align}
\label{struct3pt}
A_3=C_{\bar h_1,\bar h_2,\bar h_3}^{h_1,h_2,h_3}&(w_1-w_2)^{1-h_1-h_2+h_3}(w_2-w_3)^{h_1-h_2-h_3}(w_3-w_1)^{1-h_1+h_2-h_3}\notag\\
 &(\bar w_1-\bar w_2)^{1-\bar h_1-\bar h_2+\bar h_3}(\bar w_2-\bar w_3)^{\bar h_1-\bar h_2-\bar h_3}(\bar w_3-\bar w_1)^{1-\bar h_1+\bar h_2-\bar h_3}\\
 & \delta\left(w_1-w_2\right)\delta\left(\bar w_1-\bar w_2\right)\delta\left(w_1-w_3\right)\delta\left(\bar w_1-\bar w_3\right).\notag
\end{align}
As a cross-check we verify that Lorentz generators annihilate the three-point structure: $\sum_i M_i^{\mu\nu}A_3=0$. Since $A_3$ is a distribution as in the previous subsection, delta function derivatives will have to be evaluated in a distributional sense. To facilitate that, as previously we \textit{define} the action of the operator $\sum_i M_i^{\mu\nu}$ on $A_3$ to have the normalization
\begin{align}
\frac{1}{g_{\bar h_1,\bar h_2,\bar h_3}^{h_1,h_2,h_3}(w_1,\bar w_1,w_2,\bar w_2,w_3,\bar w_3)}\sum_i M_i^{\mu\nu}A_3=0,
\end{align}
which formally isolates the eigenvalue of $\sum_i M_i^{\mu\nu}$ next to the delta functions. 

Pairs of delta functions have overlapping dependence on $w_1$ and $\bar w_1$ in this case,\footnote{Naturally, the overlapping dependence of the delta functions on $w_1$ and $\bar w_1$ is an arbitrary but fixed choice. Permuting particle indices produces alternative three-point structure parametrizations that are equally valid.} so that delta function derivatives are resolved in a distributional sense as follows
\begin{align}
f(...)\delta'(w_1-w_2)\delta(w_1-w_3)&=f(...)\frac{\partial_{w_1}-\partial_{w_2}+\partial_{w_3}}{2}\delta(w_1-w_2)\delta(w_1-w_3)\notag\\
&=-\delta(w_1-w_2)\delta(w_1-w_3)\frac{\partial_{w_1}-\partial_{w_2}+\partial_{w_3}}{2}f(...),\\
f(...)\delta(w_1-w_2)\delta'(w_1-w_3)&=f(...)\frac{\partial_{w_1}+\partial_{w_2}-\partial_{w_3}}{2}\delta(w_1-w_2)\delta(w_1-w_3)\notag\\
&=-\delta(w_1-w_2)\delta(w_1-w_3)\frac{\partial_{w_1}+\partial_{w_2}-\partial_{w_3}}{2}f(...),
\end{align}
and similarly for $\delta'(\bar w_1-\bar w_2)\delta(\bar w_1-\bar w_3)$ and $\delta(\bar w_1-\bar w_2)\delta'(\bar w_1-\bar w_3)$. Here the extracted differential operators are fixed such that each derivative has equal weight up to an overall sign. With this, Lorentz generators readily annihilate the three-point structure.

\paragraph{Global translation invariance constraints}$~$\\
Note that in the massless case the operators $P_i^{\mu}$ in (\ref{m0P}) act non-trivially only on the conformal weights $h_i,\bar h_i$, so that the $w_{ij},\bar w_{ij}$ dependence in $A_3$ may be equivalently transformed on the support of the ideal spanned by the delta function arguments before the momentum conservation constraint is applied. For convenience we use the following transformation
\begin{align}
\label{trafo12}
w_{12}^a w_{23}^b w_{31}^c \bar{w}_{12}^{\bar{a}}
   \bar{w}_{23}^{\bar{b}} \bar{w}_{31}^{\bar{c}}&\to w_{12}^a w_{12}^b w_{12}^c \bar{w}_{12}^{\bar{a}}
   \bar{w}_{12}^{\bar{b}} \bar{w}_{12}^{\bar{c}}=w_{12}^{a+b+c}\bar w_{12}^{\bar a+\bar b+\bar c},
\end{align}
where we performed equivalent substitutions $w_2\to w_1$ and $w_3\to w_1\to w_2$ so that $w_{23}^b\to w_{12}^b$, as well as equivalent substitutions $w_3\to w_1$ and $w_1\to w_2$ so that $w_{31}^c\to w_{12}^c$ on the support of $\delta(w_{12})\delta(w_{13})$, and analogous for the barred quantities on the support of $\delta(\bar w_{12})\delta(\bar w_{13})$. Employing the same steps, $A_3$ in (\ref{struct3pt}) equivalently reads
\begin{align}
\label{A3as12}
A_3=C_{\bar h_1,\bar h_2,\bar h_3}^{h_1,h_2,h_3} w_{12}^{2-h_1-h_2-h_3}
   \bar{w}_{12}^{2-\bar{h}_1-\bar{h}_2-\bar{h}_3} \delta\left(w_{12}\right)\delta\left(\bar w_{12}\right)\delta\left(w_{13}\right)\delta\left(\bar w_{13}\right).
\end{align}
Furthermore, since $w_{12}$ is most generally a complex number of magnitude $|w_{12}|$ and some phase $\alpha$, we use
\begin{align}
w_{12}=|w_{12}|\,e^{i\alpha}~~~,~~~\bar w_{12}=|w_{12}|\,e^{-i\alpha}\,,
\end{align}
as well as $h_j=\frac{1+i\lambda_j+J_j}{2},\,\bar h_j=\frac{1+i\lambda_j-J_j}{2}$, to write $A_3$ in (\ref{A3as12}) as
\begin{align}
\label{A3phase}
A_3=C_{\bar h_1,\bar h_2,\bar h_3}^{h_1,h_2,h_3}  \left| w_{12}\right| ^{1-i (\lambda _1+ \lambda _2+ \lambda _3)} e^{-i \alpha  \left(J_1+J_2+J_3\right)}\delta\left(w_{12}\right)\delta\left(\bar w_{12}\right)\delta\left(w_{13}\right)\delta\left(\bar w_{13}\right).
\end{align}
Note that due to the positive real part in the exponent of $\left| w_{12}\right| ^{1-i (\lambda _1+ \lambda _2+ \lambda _3)}$, we have $A_3\to0$ as $w_{12}\to0$ is localized on the support of the delta function $\delta(w_{12})$. Therefore, for any helicity values $J_1,J_2,J_3$, the massless three-point structure is in general constrained to vanish when the distribution is evaluated. This agrees with the expectation that a massless three-point amplitude in real kinematics must vanish.

However, note that the application of the total momentum operator $\sum_{j} \epsilon_j P_j^{\mu}A_3$ (where $\epsilon_i=-\epsilon_j=-\epsilon_k=\pm1$ with $i,j,k\in\{1,2,3\}$ all different) changes the distribution. Making use of equivalent substitutions $w_2\to w_1,\,\bar w_2\to \bar w_1\,,w_3\to w_1\,,\bar w_3\to \bar w_1$ as above, the momentum conservation constraint simplifies to
\begin{align}
\label{Pzero3pt}
\left(\epsilon_1C_{\bar h_1+\frac{1}{2},\bar h_2,\bar h_3}^{h_1+\frac{1}{2},h_2,h_3}+\epsilon_2C_{\bar h_1,\bar h_2+\frac{1}{2},\bar h_3}^{h_1,h_2+\frac{1}{2},h_3}+\epsilon_3C_{\bar h_1,\bar h_2,\bar h_3+\frac{1}{2}}^{h_1,h_2,h_3+\frac{1}{2}}\right)\frac{q_1^\mu}{|w_{12}|}\frac{A_3}{C_{\bar h_1,\bar h_2,\bar h_3}^{h_1,h_2,h_3}}=0.
\end{align}
On the other hand, the action of the total momentum squared $(\sum_{j} \epsilon_j P_j)^{2}A_3=0$ leads to
\begin{align}
\label{Pzero3ptPP}
\left(\epsilon_1 \epsilon_2 C_{\bar h_1+\frac{1}{2},\bar h_2+\frac{1}{2},\bar h_3}^{h_1+\frac{1}{2},h_2+\frac{1}{2},h_3}+\epsilon_2 \epsilon_3 C_{\bar h_1,\bar h_2+\frac{1}{2},\bar h_3+\frac{1}{2}}^{h_1,h_2+\frac{1}{2},h_3+\frac{1}{2}}+\epsilon_3 \epsilon_1 C_{\bar h_1+\frac{1}{2},\bar h_2,\bar h_3+\frac{1}{2}}^{h_1+\frac{1}{2},h_2,h_3+\frac{1}{2}}\right)\frac{A_3}{C_{\bar h_1,\bar h_2,\bar h_3}^{h_1,h_2,h_3}}=0.
\end{align}
For all other structures the total momentum squared operator annihilation constraint is redundant, in this case however it is satisfied subtly differently. Note that, since for any values of spins of the three particles we have $A_3\sim|w_{12}|^{1-i\sum_j\lambda_j}$, the constraint (\ref{Pzero3ptPP}) is automatically satisfied by $w_{12}=0$. On the other hand, evaluating the distribution in (\ref{Pzero3pt}) the overall power of $|w_{12}|$ does not possess a positive real part, so that (despite the fact that $A_3$ itself vanishes) it leads to a non-trivial constraint on the three-point function coefficient:\footnote{Once again, we emphasize that this symmetry relation was first found by Strominger \textit{et al.} with regard to OPE coefficients and, in essence, was made public by Strominger in the talk \cite{Strominger:NEStringMeeting2019} earlier this year. Our independent calculation agrees with their finding. The content of the talk given by Strominger has since appeared in print \cite{Pate:2019lpp} with more details and should be considered as the original source of this symmetry relation.}
\begin{align}
\label{Pzero3ptC}
\epsilon_1C_{\bar h_1+\frac{1}{2},\bar h_2,\bar h_3}^{h_1+\frac{1}{2},h_2,h_3}+\epsilon_2C_{\bar h_1,\bar h_2+\frac{1}{2},\bar h_3}^{h_1,h_2+\frac{1}{2},h_3}+\epsilon_3C_{\bar h_1,\bar h_2,\bar h_3+\frac{1}{2}}^{h_1,h_2,h_3+\frac{1}{2}}=0.
\end{align}
We will see that this constraint is properly satisfied, e.g., by the gluon three-point function coefficient derived in appendix \ref{sec:gluon3pt}.

\subsection{Four-point structure}
\paragraph{Maximal minor determinant constraints}$~$\\
For $n=4$ the matrix $Q=(q_1^\mu q_2^\mu q_3^\mu q_4^\mu)$ is square, so that its only maximal minor is $Q$ itself. Demanding the vanishing of the determinant of $Q$ and fully reducing this constraint leads to several branches of solutions such that different points become degenerate ($w_i=w_j$).  However, these collinear configurations describe special kinematics, while for generic non-degenerate kinematics there exists only exactly one solution that can be summarized by the constraint
\begin{align}
z=\bar z,
\end{align}
where we have defined the cross-ratios
\begin{align}
\label{zzbar}
z\equiv\frac{(w_1-w_2)(w_3-w_4)}{(w_1-w_3)(w_2-w_4)}~~~,~~~\bar z\equiv\frac{(\bar w_1-\bar w_2)(\bar w_3-\bar w_4)}{(\bar w_1-\bar w_3)(\bar w_2-\bar w_4)}.
\end{align}
Intuitively, this can be understood as the momentum conservation constraint forcing a fourth momentum to point to the same celestial circle the other three momenta happen to point to.\footnote{As was discussed in \cite{Pasterski:2017ylz} while considering the gluon amplitude example.} Therefore, we conclude that a generic massless four-point structure on the celestial sphere must be proportional to a delta function imposing the above constraint:
\begin{align}
A_4&=\delta\left(i\bar z - i z\right)\cdot g_{\bar h_1,\bar h_2,\bar h_3,\bar h_4}^{h_1,h_2,h_3,h_4}(w_1,\bar w_1,w_2,\bar w_2,w_3,\bar w_3,w_4,\bar w_4)
\end{align}
where $g$ at this point is an arbitrary function of its arguments.

\paragraph{Lorentz invariance constraints}$~$\\
Since the delta function only depends on conformal cross-ratios $z,\bar z$, in this case all delta function derivatives cancel out of the action of Lorentz generators $\sum_i M_i^{\mu\nu}$. Thus the delta function becomes a spectator and the four-point structure can essentially be treated as a function instead of a distribution. 

Considering that the Lorentz part of Poincar\'e invariance implies the same constraints as the usual conformal covariance of the four-point structure on the celestial sphere, a convenient way to write a generic conformally covariant four-point structure of primary fields is by using the conventional pre-factor (see, e.g., \cite{Osborn:2012vt})
\begin{align}
 F_{n=4}\equiv\frac{\left(\frac{w_1-w_4}{w_1-w_3}\right)^{h_3-h_4}
   \left(\frac{w_2-w_4}{w_1-w_4}\right)^{h_1-h_2}}{\left(w_1-w_2\right)^{h_1+h_2} \left(w_3-w_4\right)^{h_3+h_4}} 
   \frac{\left(\frac{\bar{w}_1-\bar{w}_4}{\bar{w}_1-\bar{w}_3}\right)^{\bar{h}_3-\bar{h}_4}
   \left(\frac{\bar{w}_2-\bar{w}_4}{\bar{w}_1-\bar{w}_4}\right)^{\bar{h}_1-\bar{h}_2}}{\left(\bar{w}_1-\bar{w}_2\right)^{\bar{h}_1+\bar{h}_2}\left(\bar{w}_3-\bar{w}_4\right)^{\bar{h}_3+\bar{h}_4}},
\end{align}
so that
\begin{align}
A_4&=F_{n=4}\cdot \delta\left(i\bar z - i z\right) f_{\bar h_1,\bar h_2,\bar h_3,\bar h_4}^{h_1,h_2,h_3,h_4}(z ,\bar z),
\end{align}
where $f_{\bar h_1,\bar h_2,\bar h_3,\bar h_4}^{h_1,h_2,h_3,h_4}(z ,\bar z)$, similarly to $\delta\left(i\bar z - i z\right)$, is a function of the cross ratios $z,\bar z$.

It is then straightforward to verify that the above structure indeed is properly annihilated by all Lorentz generators $\sum_i M_i^{\mu\nu}$ (in this case the normalization is irrelevant). Since we are making use of the factor $F_{n=4}$ in our parametrization, M\"obius transformations lead to correct primary field transformation weights per definition.

\paragraph{Global translation invariance constraints}$~$\\
As a next step, we proceed to solve the equations $\sum_{j} \epsilon_j P_j^{\mu}A_4=0$ to find further constraints on $f_{\bar h_1,\bar h_2,\bar h_3,\bar h_4}^{h_1,h_2,h_3,h_4}(z ,\bar z)$. We recall the following cross ratio regions for different incoming and outgoing particle configurations, which always arise in massless 4-pt celestial amplitudes due to the presence of total momentum conservation delta functions in Minkowski space amplitudes regardless of any other amplitude features (see, e.g., example in \cite{Pasterski:2017ylz}):
\begin{align}
\epsilon_1=&\epsilon_2=-\epsilon_3=-\epsilon_4~~~\Rightarrow~~~1<z,\\
\epsilon_1=&\epsilon_3=-\epsilon_2=-\epsilon_4~~~\Rightarrow~~~0<z<1,\\
\epsilon_1=&\epsilon_4=-\epsilon_2=-\epsilon_3~~~\Rightarrow~~~z<0.
\end{align}
Additionally, making use of the cross ratio relation $z=\bar z$, the momentum conservation equations simplify and equivalently reduce to\footnote{A quick way to arrive at the simplification in terms of conformal cross ratios is to employ conformal invariance of $f$ and fix three of the four points to the particular values
$w_1=\bar w_1=0,~w_2=z,\,\bar w_2=\bar z,~w_3=\bar w_3=1,~w_4=\bar w_4=\infty$.}
\begin{align}
f_{\bar h_1,\bar h_2,\bar h_3+\frac{1}{2},\bar h_4}^{h_1,h_2,h_3+\frac{1}{2},h_4}(z ,\bar z)&=\underbrace{\left(\epsilon_3\epsilon_4\text{sgn}(z-1)\right)}_{=1}\frac{1}{|z-1|}f_{\bar h_1,\bar h_2,\bar h_3,\bar h_4+\frac{1}{2}}^{h_1,h_2,h_3,h_4+\frac{1}{2}}(z ,\bar z),\notag\\
\label{A4fconstr}
f_{\bar h_1,\bar h_2+\frac{1}{2},\bar h_3,\bar h_4}^{h_1,h_2+\frac{1}{2},h_3,h_4}(z ,\bar z)&=\underbrace{\left(\epsilon_2\epsilon_4\text{sgn}\big(z(1-z)\big)\right)}_{=1}\frac{1}{|z-1|}f_{\bar h_1,\bar h_2,\bar h_3,\bar h_4+\frac{1}{2}}^{h_1,h_2,h_3,h_4+\frac{1}{2}}(z ,\bar z),\\
f_{\bar h_1+\frac{1}{2},\bar h_2,\bar h_3,\bar h_4}^{h_1+\frac{1}{2},h_2,h_3,h_4}(z ,\bar z)&=\underbrace{\left(-\epsilon_1\epsilon_4\text{sgn}\big(z\big)\right)}_{=1}f_{\bar h_1,\bar h_2,\bar h_3,\bar h_4+\frac{1}{2}}^{h_1,h_2,h_3,h_4+\frac{1}{2}}(z ,\bar z).\notag
\end{align}
As expected, due to the reduced rank of coefficient matrix $Q$, the four momentum conservation constraints lead to only three linearly independent constraints (\ref{A4fconstr}). 

Without loss of generality, the function $f_{\bar h_1,\bar h_2,\bar h_3,\bar h_4}^{h_1,h_2,h_3,h_4}(z ,\bar z)$ can be parametrized as
\begin{align}
\label{A4f}
f_{\bar h_1,\bar h_2,\bar h_3,\bar h_4}^{h_1,h_2,h_3,h_4}(z ,\bar z)&=(z-1)^{\frac{h_1-h_2-h_3+h_4}{2}}(\bar z-1)^{\frac{\bar h_1-\bar h_2-\bar h_3+\bar h_4}{2}}\tilde f_{\Delta_1,\Delta_2,\Delta_3,\Delta_4}^{J_1,J_2,J_3,J_4}(z ,\bar z),
\end{align}
where $\tilde f_{\Delta_1,\Delta_2,\Delta_3,\Delta_4}^{J_1,J_2,J_3,J_4}(z ,\bar z)$ is a new generic function of its arguments.\footnote{Generality is not lost since all arguments of $\tilde f_{\Delta_1,\Delta_2,\Delta_3,\Delta_4}^{J_1,J_2,J_3,J_4}(z ,\bar z)$ are so far unconstrained, so that the factor in front may be absorbed to recover $f_{\bar h_1,\bar h_2,\bar h_3,\bar h_4}^{h_1,h_2,h_3,h_4}(z ,\bar z)$.}

Making use of (\ref{A4f}) in (\ref{A4fconstr}) leads to the following simpler constraints on $\tilde f$:
\begin{align}
\tilde f_{\Delta_1,\Delta_2,\Delta_3+1,\Delta_4}^{J_1,J_2,J_3,J_4}(z ,\bar z)&=\tilde f_{\Delta_1,\Delta_2,\Delta_3,\Delta_4+1}^{J_1,J_2,J_3,J_4}(z ,\bar z),\notag\\
\label{A4fconstr2}
\tilde f_{\Delta_1,\Delta_2+1,\Delta_3,\Delta_4}^{J_1,J_2,J_3,J_4}(z ,\bar z)&=\tilde f_{\Delta_1,\Delta_2,\Delta_3,\Delta_4+1}^{J_1,J_2,J_3,J_4}(z ,\bar z),\\
\tilde f_{\Delta_1+1,\Delta_2,\Delta_3,\Delta_4}^{J_1,J_2,J_3,J_4}(z ,\bar z)&=\tilde f_{\Delta_1,\Delta_2,\Delta_3,\Delta_4+1}^{J_1,J_2,J_3,J_4}(z ,\bar z).\notag
\end{align}
In appendix \ref{A4shifts} we solve for the generic set of functions $\tilde f$ satisfying the constraints (\ref{A4fconstr2}).

Collecting everything together, Poincar\'e invariance ensures that the celestial four point structure for massless particles can always be written as
\begin{align}
\label{zeromass4ptA}
\begin{split}
A_4=&\frac{\left(\frac{w_1-w_4}{w_1-w_3}\right)^{h_3-h_4}
   \left(\frac{w_2-w_4}{w_1-w_4}\right)^{h_1-h_2}}{\left(w_1-w_2\right)^{h_1+h_2} \left(w_3-w_4\right)^{h_3+h_4}} 
   \frac{\left(\frac{\bar{w}_1-\bar{w}_4}{\bar{w}_1-\bar{w}_3}\right)^{\bar{h}_3-\bar{h}_4}
   \left(\frac{\bar{w}_2-\bar{w}_4}{\bar{w}_1-\bar{w}_4}\right)^{\bar{h}_1-\bar{h}_2}}{\left(\bar{w}_1-\bar{w}_2\right)^{\bar{h}_1+\bar{h}_2}\left(\bar{w}_3-\bar{w}_4\right)^{\bar{h}_3+\bar{h}_4}}\\
& \delta\left(i\bar z - i z\right)(z-1)^{\frac{h_1-h_2-h_3+h_4}{2}}(\bar z-1)^{\frac{\bar h_1-\bar h_2-\bar h_3+\bar h_4}{2}}\tilde f_{\Delta_1,\Delta_2,\Delta_3,\Delta_4}^{J_1,J_2,J_3,J_4}(z ,\bar z),
\end{split}
\end{align}
where, as explained in appendix \ref{A4shifts}, the functional $\Delta_i$ dependence in $\tilde f$ must either appear as the combination $\sum_{i=1}^4\Delta_i$, or else be periodic of period $1$. Note that despite the $z=\bar z$ constraint, we maintain a complex notation to emphasize the single-valuedness of the result. Especially the momentum generator constraints have led to special features that are not familiar from usual CFT.

Recall that the inverse transform from celestial amplitudes back to Minkowski space amplitudes involves Mellin-like integration over $-\infty<\lambda<\infty$ for each conformal dimension $\Delta=1+i\lambda$ of external particles \cite{Pasterski:2017kqt}. Convergence of this inverse integral transform requires the $\Delta$ dependence of celestial amplitudes to be analytic in the vicinity of the complex line of continuous series representation $\Delta=1+i\lambda$, and to uniformly tend to zero as $\lambda\to\pm\infty$. Note however, that a non-trivial holomorphic function of finite order\footnote{Finite order functions are sufficiently well behaved, which rules out piculiarities such as doubly-exponential functions. Since on physical grounds we expect to obtain the inverse integral transform from integration contour deformation, collecting finite order residues and discontinuities, it is natural to impose this restriction.} that is periodic
\begin{align}
\label{boundedness}
f(\Delta+1)=f(\Delta),
\end{align}
must be unbounded somewhere on any complex vertical line in $\Delta$. This can be seen as follows. Assuming the function was bounded along one complex vertical line, by periodicity the same line would repeat, so that the function would be bounded on the boundary of an infinite strip of finite width. In such a case, the so called Phragm\'en-Lindel\"of method in complex analysis shows that the maximum modulus principle can be applied, which leads to the conclusion that the function must be bounded not only on the boundaries but on the whole infinite strip, and, by periodicity, on the whole complex plane. It is a known fact that a function that is bounded on the whole complex plane must be constant. Therefore, to avoid being constant, a non-trivial holomorphic periodic function as described above must be unbounded on all possible vertical complex lines, which represents a non-analyticity and therefore spoils convergence properties of the inverse integral transform if left unremedied.

Taking the above considerations into account, a function $\tilde f$ in (\ref{zeromass4ptA}) involving non-constant periodic dependence on conformal dimensions $\Delta$ should simultaneously have carefully fine-tuned non-periodic dependence on $\sum_i\Delta_i$ to remedy the unboundedness on the complex vertical line arising from the periodic dependence, if possible. If no periodic dependence is present, then the $\sum_i\Delta_i$ dependence may be arbitrary.

\section{Poincar\'e constraints on 2,3 and 4-point structures with massive scalars}
\label{sec:massive}
Since the operators (\ref{PmS}) involve different types of derivatives and shifts, the action of the total momentum generator cannot be thought of as an overdetermined linear set of equations. Therefore, in the massive case there are no minor determinant constraints and all the $n$-point structures are regular functions instead of distributions. With this, the Lorentz symmetry imposes familiar conformal symmetry constraints on the celestial sphere, such that conformally covariant two-, three- and four-point structures can be written as \cite{DiFrancesco:1997nk,Osborn:2012vt}
\begin{align}
\label{A2m}
A_2=&C^{h_1,h_2}_{\bar h_1,\bar h_2}\frac{\hat\delta(h_1-h_2)\hat\delta(\bar h_1-\bar h_2)}{w_{12}^{h_1+h_2}\bar w_{12}^{\bar h_1+\bar h_2}},\\
\label{A3m}
A_3=&C_{\bar h_1,\bar h_2,\bar h_3}^{h_1,h_2,h_3}w_{12}^{h_3-h_1-h_2}w_{23}^{h_1-h_2-h_3}w_{31}^{h_2-h_1-h_3}
 \bar w_{12}^{\bar h_3-\bar h_1-\bar h_2}\bar w_{23}^{\bar h_1-\bar h_2-\bar h_3}\bar w_{31}^{\bar h_2-\bar h_1-\bar h_3},\\
\label{A4m}
A_4=&\frac{\left(\frac{w_{14}}{w_{13}}\right)^{h_3-h_4}
   \left(\frac{w_{24}}{w_{14}}\right)^{h_1-h_2}}{w_{12}^{h_1+h_2} w_{34}^{h_3+h_4}} 
   \frac{\left(\frac{\bar{w}_{14}}{\bar{w}_{13}}\right)^{\bar{h}_3-\bar{h}_4}
   \left(\frac{\bar{w}_{24}}{\bar{w}_{14}}\right)^{\bar{h}_1-\bar{h}_2}}{\bar{w}_{12}^{\bar{h}_1+\bar{h}_2}\bar{w}_{34}^{\bar{h}_3+\bar{h}_4}}f_{\bar h_1,\bar h_2,\bar h_3,\bar h_4}^{h_1,h_2,h_3,h_4}(z ,\bar z),
\end{align}
with conformal cross ratios $z,\bar z$ defined in (\ref{zzbar}), and the finite-valued delta function $\hat \delta(x)$ as defined in (\ref{findelta}). Once again, from current considerations it is not clear whether the two-point structure may diverge when the conformal weights are equal; in such case we can use a combination of Dirac and Kronecker deltas instead
\begin{align}
\hat\delta(h_1-h_2)\hat\delta(\bar h_1-\bar h_2)~~\to~~\delta(\Delta_1-\Delta_2)\delta_{J_1,J_2},
\end{align}
again with no change to further conclusions regarding constraints on the two-point function coefficient.
We expect to find further constraints on coefficients $C$ and the function of cross ratio $f$ from global translation invariance.

\subsection{Two-point structure}
The equations $\sum_{j} \epsilon_j P_j^{\mu}A_2^{m_1,m_2}=0$ for two massive scalar two-point structure, with momentum operators (\ref{PmS}) and $A_2$ in (\ref{A2m}) such that $h_j=\bar h_j=\frac{\Delta_j}{2}$, lead to two independent conditions each proportional to $\hat\delta(\frac{\Delta_1-\Delta_2\pm 1}{2})$ respectively. On the support of $\hat\delta(\frac{\Delta_1-\Delta_2- 1}{2})$, the vanishing condition reduces to
\begin{align}
(\Delta-1)m_1 C^{\frac{\Delta}{2},\frac{\Delta}{2}}_{\frac{\Delta}{2},\frac{\Delta}{2}} -\Delta m_2 C^{\frac{\Delta+1}{2},\frac{\Delta+1}{2}}_{\frac{\Delta+1}{2},\frac{\Delta+1}{2}}=0,
\end{align}
where we abbreviate $\Delta_2=\Delta,~\Delta_1=1+\Delta$. Similarly, on the support of $\hat\delta(\frac{\Delta_1-\Delta_2+ 1}{2})$, the vanishing condition reduces to
\begin{align}
(\Delta-1)m_2 C^{\frac{\Delta}{2},\frac{\Delta}{2}}_{\frac{\Delta}{2},\frac{\Delta}{2}} -\Delta m_1 C^{\frac{\Delta+1}{2},\frac{\Delta+1}{2}}_{\frac{\Delta+1}{2},\frac{\Delta+1}{2}}=0,
\end{align}
where now we abbreviate $\Delta_1=\Delta,~\Delta_2=1+\Delta$. It is clear that both resulting vanishing conditions can be reconciled only if we demand
\begin{align}
m_1=m_2\equiv m,
\end{align}
in which case the constraint is solved by
\begin{align}
\label{massive2ptC}
C^{\frac{\Delta}{2},\frac{\Delta}{2}}_{\frac{\Delta}{2},\frac{\Delta}{2}}=\frac{c_{\Delta}}{\Delta-1}~~~\Rightarrow~~~C^{\frac{\Delta_1}{2},\frac{\Delta_2}{2}}_{\frac{\Delta_1}{2},\frac{\Delta_2}{2}}=\frac{c_{\frac{\Delta_1+\Delta_2}{2}}}{\frac{\Delta_1+\Delta_2}{2}-1},
\end{align}
with the arbitrary unknown $c_x$, periodic under shifts in $x$ of period $1$: $c_{x+1}=c_{x}$. Note that, to avoid convergence issues with the inverse integral transform from the celestial sphere to Minkowski space, as discussed around (\ref{boundedness}), $c_{x}$ better be trivially periodic $c_{x}=const.$. However, note that the inverse transform along the vertical line $\Delta=1+i\lambda$ with $-\infty<\lambda<\infty$ exactly hits the pole in (\ref{massive2ptC}), such that the inverse transform fails and in Minkowski space there exists no counterpart for this two-point function.

As in the completely massless case, a two-point structure, e.g., for $m_1=0,~m_2=m>0$ does not exist since the two-point function coefficient is then constrained to vanish.

\subsection{Three-point structure}
In the three-point case we can consider configurations involving one, two, or three massive scalars. We study them in increasing order.

\paragraph{Two massless, one massive:} To satisfy the equations $\sum_{j} \epsilon_j P_j^{\mu}A_3^{0,0,m}=0$, we calculate the operator action, divide out the common dependence on $w_j,\bar w_j$ in each term and demand that the coefficient of each resulting monomial in $w_j,\bar w_j$ vanishes separately. With the third particle being a scalar of mass $m$, the vanishing of all coefficients can only be non-trivially satisfied when the two massless particles have the same spin
\begin{align}
\label{spin12eq}
h_1-\bar h_1=J_1\equiv J\equiv J_2=h_2-\bar h_2,
\end{align}
and  lead to the following conditions on the three-point function coefficient $C$:
\begin{align}
0=&\left((h_1-h_2)^2-\left(\frac{\Delta_3-1}{2}\right)^2\right)C_{\bar h_1,\bar h_2,\frac{\Delta_3-1}{2}}^{h_1,h_2,\frac{\Delta_3-1}{2}}+\Delta_3(\Delta_3-1)C_{\bar h_1,\bar h_2,\frac{\Delta_3+1}{2}}^{h_1,h_2,\frac{\Delta_3+1}{2}},\notag\\
\label{C00m1}
0=&4 \epsilon_2 (\Delta_3-1) C_{\bar h_1,\bar h_2+\frac{1}{2},\frac{\Delta_3}{2}}^{h_1,h_2+\frac{1}{2},\frac{\Delta_3}{2}}+m \epsilon_3 (\Delta_3-1-2h_1+2h_2) C_{\bar h_1,\bar h_2,\frac{\Delta_3-1}{2}}^{h_1,h_2,\frac{\Delta_3-1}{2}},\\
0=&4 \epsilon_1 (\Delta_3-1) C_{\bar h_1+\frac{1}{2},\bar h_2,\frac{\Delta_3}{2}}^{h_1+\frac{1}{2},h_2,\frac{\Delta_3}{2}}+m \epsilon_3 (\Delta_3-1+2h_1-2h_2) C_{\bar h_1,\bar h_2,\frac{\Delta_3-1}{2}}^{h_1,h_2,\frac{\Delta_3-1}{2}}.\notag
\end{align}
These constraints are properly satisfied, e.g., by the three-point function coefficient for the case of two massless scalars and one massive scalar obtained in \cite{Lam:2017ofc}. Without loss of generality (and making use of $h_1-h_2=\bar h_1-\bar h_2$) we may rescale the three-point function coefficient as
\begin{align}
\label{resC00m}
C_{\bar h_1,\bar h_2,\frac{\Delta_3}{2}}^{h_1,h_2,\frac{\Delta_3}{2}}=\frac{\left(\frac{m }{2}\right)^{\bar{h}_1+\bar{h}_2+h_1+h_2} \Gamma \left(h_1-h_2+\frac{\Delta _3}{2}\right) \Gamma \left(\frac{\Delta _3}{2}-h_1+h_2\right)}{\Gamma \left(\Delta _3\right)}C_{\Delta_1,\Delta_2,\Delta_3}^{J},
\end{align}
where for now $C_{\Delta_1,\Delta_2,\Delta_3}^{J}$ is a new arbitrary function of its arguments.
Keeping in mind that kinematically the massive leg must always be outgoing while the two massless legs are incoming or vice versa $\epsilon_1=\epsilon_2=-\epsilon_3$, we use (\ref{resC00m}) in (\ref{C00m1}), which reduce to the simpler relations
\begin{align}
\label{C00m2}
C_{\Delta_1,\Delta_2,\Delta_3-1}^{J}&=C_{\Delta_1,\Delta_2,\Delta_3+1}^{J},\\
\label{C00m3}
C_{\Delta_1+1,\Delta_2,\Delta_3}^{J}&=C_{\Delta_1,\Delta_2+1,\Delta_3}^{J},\\
\label{C00m4}
C_{\Delta_1+1,\Delta_2,\Delta_3}^{J}&=C_{\Delta_1,\Delta_2,\Delta_3-1}^{J}.
\end{align}
Due to (\ref{C00m2}), the functional dependence of $C$ on $\Delta_3$ must be periodic with period $2$. From (\ref{C00m3}), analogously to steps in appendix \ref{A4shifts}, $C$ is constrained further such that all functional dependence on $\Delta_1$ or $\Delta_2$ must be periodic with period $1$ if the dependence does not appear in combination $\Delta_1+\Delta_2$. Finally, (\ref{C00m4}) further constrains the three-point function coefficient, similarly to steps in appendix \ref{A4shifts}, such that any functional dependence on $\Delta_i$ must be periodic with period $1$, unless if it appears in the combination $\Delta_1+\Delta_2+\Delta_3$, in which case the periodicity must be of period $2$.

Assuming that $C$ is a holomorphic function of finite order, as discussed around (\ref{boundedness}), non-trivial periodicity in function $C$ causes it to become unbounded on vertical lines in the complex plane. While the gamma function ratio in (\ref{resC00m}) does tend to zero when the $\Delta_i$ approach complex infinity, e.g. entire periodic functions $C=C(e^{2\pi i\Delta})$ generically feature arbitrarily high modes $e^{2\pi i n\Delta}$ with $n\in\mathbb{Z}$ in their Laurent expansion, most of which diverge stronger than the suppressing effect of the gamma function ratio. To avoid convergence issues in the inverse transform of celestial amplitudes to Minkowski space, function $C$ better be constant in conformal dimensions
\begin{align}
C_{\Delta_1,\Delta_2,\Delta_3}^{J}=C^J\,.
\end{align}
The three-point function coefficient for the case of two massless scalars and one massive scalar \cite{Lam:2017ofc} is thus precisely found in (\ref{resC00m}) purely from symmetry in the simplest case 
$C_{\Delta_1,\Delta_2,\Delta_3}^{J}=const.\,,~J=0\,,~\epsilon_1=\epsilon_2=-\epsilon_3$.

\paragraph{One massless, two massive:} The equations $\sum_{j} \epsilon_j P_j^{\mu}A_3^{m_1,m_2,0}=0$, with the first and second particles being scalars of mass $m_1,m_2$ respectively, lead to different constraints depending on whether the massless particle is a scalar or has spin. In the case when the massless particle is a scalar
\begin{align}
h_3=\bar h_3=\frac{\Delta_3}{2},
\end{align}
the resulting constraints read
\begin{align}
0=&\frac{2 \left(\Delta _1-1\right) m_2 \epsilon
   _2 C_{\frac{\Delta _1}{2},\frac{\Delta _2-1}{2} ,\frac{\Delta _3}{2}}^{\frac{\Delta _1}{2},\frac{\Delta _2-1}{2}
   ,\frac{\Delta _3}{2}}}{(\Delta _2-1)\left(\Delta _1-\Delta _2+\Delta _3-1\right)}+\frac{4
   \Delta _1 m_1 \epsilon _1 C_{\frac{\Delta _1+1}{2}
   ,\frac{\Delta _2}{2},\frac{\Delta
   _3}{2}}^{\frac{\Delta _1+1}{2} ,\frac{\Delta
   _2}{2},\frac{\Delta _3}{2}}}{(\Delta _1+\Delta _2-\Delta _3-1)\left(\Delta _1-\Delta _2+\Delta _3-1\right)}-\frac{ m_1 \epsilon _1
   C_{\frac{\Delta _1-1}{2} ,\frac{\Delta _2}{2},\frac{\Delta _3}{2}}^{\frac{\Delta _1-1}{2},\frac{\Delta _2}{2},\frac{\Delta
   _3}{2}}}{\Delta _1-1},\notag\\
	0=&\frac{2 \left(\Delta _2-1\right) m_1 \epsilon _1 C_{\frac{\Delta _1-1}{2} ,\frac{\Delta _2}{2},\frac{\Delta _3}{2}}^{\frac{\Delta _1-1}{2}
   ,\frac{\Delta _2}{2},\frac{\Delta _3}{2}}}{(\Delta
   _1-1)\left(\Delta _2-\Delta _1+\Delta _3-1\right)}+\frac{4 \Delta _2 m_2
   \epsilon _2 C_{\frac{\Delta _1}{2},\frac{\Delta _2+1}{2} ,\frac{\Delta _3}{2}}^{\frac{\Delta _1}{2},\frac{\Delta _2+1}{2}
   ,\frac{\Delta _3}{2}}}{(\Delta _1+\Delta _2-\Delta
   _3-1)\left(\Delta _2-\Delta _1+\Delta _3-1\right)}-\frac{ m_2 \epsilon _2
   C_{\frac{\Delta _1}{2},\frac{\Delta _2-1}{2} ,\frac{\Delta _3}{2}}^{\frac{\Delta _1}{2},\frac{\Delta _2-1}{2} ,\frac{\Delta _3}{2}}}{\Delta _2-1},\notag\\
	\label{rec3ptC0m1m2}
	0=&\frac{4 \epsilon _3 C_{\frac{\Delta _1}{2},\frac{\Delta _2}{2},\frac{\Delta
   _3+1}{2}}^{\frac{\Delta _1}{2},\frac{\Delta
   _2}{2},\frac{\Delta _3+1}{2}}}{\left(\Delta _2-\Delta _1+\Delta _3-1\right)}+\frac{\left(\Delta _1-\Delta
   _2+\Delta _3-1\right) m_1 \epsilon _1 C_{\frac{\Delta _1-1}{2} ,\frac{\Delta _2}{2},\frac{\Delta _3}{2}}^{\frac{\Delta _1-1}{2}
   ,\frac{\Delta _2}{2},\frac{\Delta _3}{2}}}{(\Delta _1-1)\left(\Delta _2-\Delta _1+\Delta _3-1\right)}
	+\frac{ m_2 \epsilon _2
   C_{\frac{\Delta _1}{2},\frac{\Delta _2-1}{2} ,\frac{\Delta
   _3}{2}}^{\frac{\Delta _1}{2},\frac{\Delta _2-1}{2} ,\frac{\Delta _3}{2}}}{(\Delta _2-1)}.
	\end{align}
	There are terms with shifts in different parameters in each of the three equations, which makes finding a general solution highly non-trivial. It is likely that possible solutions may involve ${}_2F_1$ hypergeometric functions, since these functions satisfy similar three-term recurrence relations with integer shifts.
	
	On the other hand, in case when the massless particle has non-zero spin
	\begin{align}
h_3- \bar h_3=J_3\neq 0,
\end{align}
	the overall constraints are given by
	\begin{align}
0=&\left(\Delta _1+\Delta _2-2 h_3-1\right) \left(\Delta _1+\Delta
   _2-2 \bar{h}_3-1\right) C_{\frac{\Delta _1-1}{2} ,\frac{\Delta
   _2}{2},\bar{h}_3}^{\frac{\Delta _1-1}{2} ,\frac{\Delta
   _2}{2},h_3}+4 \left(\Delta _1-1\right) \Delta _1 C_{\frac{\Delta _1+1}{2}
   ,\frac{\Delta _2}{2},\bar{h}_3}^{\frac{\Delta _1+1}{2}
   ,\frac{\Delta _2}{2},h_3},\\
	0=& m_2 \epsilon _2 \Delta _1 C_{\frac{\Delta _1+1}{2},\frac{\Delta _2}{2}
   ,\bar{h}_3}^{\frac{\Delta _1+1}{2},\frac{\Delta _2}{2}
   ,h_3}- m_1 \epsilon _1 \Delta _2
   C_{\frac{\Delta _1}{2} ,\frac{\Delta
   _2+1}{2},\bar{h}_3}^{\frac{\Delta _1}{2} ,\frac{\Delta
   _2+1}{2},h_3},\\
	0=&m_1 \epsilon _1  \Delta _1 C_{\frac{\Delta _1+1}{2}
   ,\frac{\Delta _2}{2},\bar{h}_3}^{\frac{\Delta _1+1}{2}
   ,\frac{\Delta _2}{2},h_3}-  m_2 \epsilon _2\Delta _2
   C_{\frac{\Delta _1}{2},\frac{\Delta _2+1}{2} ,\bar{h}_3}^{\frac{\Delta _1}{2},\frac{\Delta _2+1}{2} ,h_3},\\
	0=&2 \left(\Delta _1-1\right) \epsilon _3 C_{\frac{\Delta _1}{2},\frac{\Delta
   _2}{2},\bar{h}_3+\frac{1}{2}}^{\frac{\Delta _1}{2},\frac{\Delta
   _2}{2},h_3+\frac{1}{2}}+m_1 \epsilon _1 \left(\bar{h}_3+h_3-1\right)
   C_{\frac{\Delta _1-1}{2},\frac{\Delta
   _2}{2},\bar{h}_3}^{\frac{\Delta _1-1}{2} ,\frac{\Delta
   _2}{2},h_3}.
	\end{align}
Note that the second and third equations can only be reconciled if we demand
	\begin{align}
	\label{mass12eq}
m_1=m_2\equiv m.
	\end{align}
Without loss of generality, the three-point function coefficient can be rescaled as
\begin{align}
C_{\frac{\Delta _1}{2} ,\frac{\Delta
   _2}{2},\bar{h}_3}^{\frac{\Delta _1}{2} ,\frac{\Delta
   _2}{2},h_3}=\left(\frac{m}{2}\right)^{\bar{h}_3+h_3}\frac{  \Gamma
   \left(h_3+\bar{h}_3-1\right) \Gamma \left(\frac{\Delta _1+\Delta _2-2 h_3}{2}
   \right) \Gamma
   \left(\frac{\Delta _1+\Delta _2-2
   \bar{h}_3}{2} \right) }{\Gamma \left(\Delta _1\right)
   \Gamma \left(\Delta _2\right)}C_{\Delta _1,\Delta
   _2,\Delta_3}^{J_3},
\end{align}
so that in terms of $C_{\Delta _1,\Delta _2,\Delta_3}^{J_3}$ the constraints reduce to
\begin{align}
\label{A3Cmm0c}
0=&C_{\Delta _1+1,\Delta _2,\Delta_3}^{J_3}+C_{\Delta _1-1,\Delta _2,\Delta_3}^{J_3},\\
0=&\epsilon_1 C_{\Delta _1+1,\Delta _2,\Delta_3}^{J_3}-\epsilon_2 C_{\Delta _1,\Delta _2+1,\Delta_3}^{J_3},\\
0=&\epsilon_1 C_{\Delta _1-1,\Delta _2,\Delta_3}^{J_3}+\epsilon_3 C_{\Delta _1,\Delta _2,\Delta_3+1}^{J_3}.
\end{align}
In this case, (\ref{A3Cmm0c}) implies that $C$ must be an anti-periodic function in $\Delta_1$ with anti-period equal to 2.\footnote{The same is of course also true for $\Delta_2$, which can be seen by employing the other two constraints.} Assuming otherwise the same function properties, the considerations discussed around (\ref{boundedness}) similarly apply, such that a non-trivial anti-periodic $C$ will be unbounded somewhere on any complex line in $\Delta_1$ stretching from $-i\infty$ to $i\infty$, causing convergence issues in the inverse integral transform of the celestial amplitude to Minkowski space. To avoid convergence issues we should take the trivial solution that does not depend on higher anti-periodic Laurent expansion modes $e^{i\frac{\pi(2n+1)}{2}\Delta_1}$ with $n\in\mathbb{Z}$ under $\Delta_1\to\Delta_1+2$. However, the trivial anti-periodic function without higher exponential modes reads
\begin{align}
C_{\Delta _1,\Delta _2,\Delta_3}^{J_3}=0,
\end{align}
since any other constant would not be anti-periodic.
	The vanishing of this amplitude can be understood from the Minkowski space perspective as well, as the limit $m_2\to m_1$ kinematically forces the massless particle momentum to tend to zero $p_3^\mu\to0$. In the strict limit this creates an unphysical kinematic configuration, such that the amplitude is ruled out.\footnote{We thank the referee for emphasizing this point.}
	
	\paragraph{Zero massless, three massive:} The equations $\sum_{j} \epsilon_j P_j^{\mu}A_3^{m_1,m_2,m_3}=0$, with all three particles being massive scalars of mass $m_1,m_2,m_3$ respectively, reduce to the following three constraints on the three-point function coefficient $C=C_{\frac{\Delta _1}{2},\frac{\Delta _2}{2},\frac{\Delta _3}{2}}^{\frac{\Delta _1}{2},\frac{\Delta _2}{2},\frac{\Delta _3}{2}}$:
		\begin{align}
\label{rec3ptCmmm}
	0=&\frac{4  \Delta _i m_i \epsilon _i  \left(-\Delta _i+\Delta _j+\Delta
   _k-1\right){}^2}{\left(\Delta _i+\Delta _j-\Delta _k-1\right) \left(\Delta _i-\Delta
   _j+\Delta _k-1\right)}e^{\partial_{\Delta _i}}C-\frac{ m_i \epsilon _i  \left(\Delta
   _i+\Delta _j+\Delta _k-3\right){}^2}{\Delta _i-1}e^{-\partial_{\Delta _i}}C\notag\\
	&-\frac{8  \left(\Delta _i-1\right)
   \Delta _j m_j \epsilon _j }{\Delta _i+\Delta _j-\Delta _k-1}e^{\partial_{\Delta _j}}C-\frac{8 
   \left(\Delta _i-1\right) \Delta _k m_k \epsilon _k }{\Delta _i-\Delta
   _j+\Delta _k-1}e^{\partial_{\Delta _k}}C,
	\end{align}
	one equation for each triplet of indices $(i,j,k)\in\{(1,2,3),(2,1,3),(3,1,2)\}$.
	As in the case of one massless and two massive scalars, the above constraints are highly non-trivial to solve in general. However, it is likely that possible solutions may involve ${}_3F_2$ hypergeometric functions, since these functions satisfy similar four-term recurrence relations with integer shifts. Additionally, the recurrence relations provide a set of conditions that necessarily have to be satisfied as a non-trivial cross-check, should the result be obtained by other means in the future.

\subsection{Four-point structure}
In the four-point case we can consider configurations involving one, two, three, or four massive scalars. We take a closer look at the first case and summarize the features of the remaining cases.

\paragraph{Three massless, one massive:} The equations $\sum_{j} \epsilon_j P_j^{\mu}A_4^{0,0,0,m}=0$, with the fourth particle being a scalar of mass $m$, lead to four different consistency conditions. One of the conditions involves only shifts in operator dimensions\footnote{Once again, a quick way to arrive at the simplification in terms of conformal cross ratios is to employ conformal invariance of $f$ and fix three of the four points to the particular values
$w_1=\bar w_1=0,~w_2=z,\,\bar w_2=\bar z,~w_3=\bar w_3=1,~w_4=\bar w_4=\infty$.}
	\begin{align}
0=&\sqrt{z \bar{z}}\left(m \epsilon _4  f_{\bar{h}_1,\bar{h}_2,\bar{h}_3,\frac{\Delta _4-1}{2}
   }^{h_1,h_2,h_3,\frac{\Delta
   _4-1}{2} }+2 \epsilon _3 
   f_{\bar{h}_1,\bar{h}_2,\bar{h}_3+\frac{1}{2},\frac{\Delta
   _4}{2}}^{h_1,h_2,h_3+\frac{1}{2},\frac{\Delta
   _4}{2}}\right)+2 \epsilon _2
   f_{\bar{h}_1,\bar{h}_2+\frac{1}{2},\bar{h}_3,\frac{\Delta
   _4}{2}}^{h_1,h_2+\frac{1}{2},h_3,\frac{\Delta
   _4}{2}}+2 \epsilon _1
   f_{\bar{h}_1+\frac{1}{2},\bar{h}_2,\bar{h}_3,\frac{\Delta
   _4}{2}}^{h_1+\frac{1}{2},h_2,h_3,\frac{\Delta
   _4}{2}},
	\end{align}
while the remaining three conditions involve shifts in operator dimensions as well as derivatives with respect to the holomorphic and anti-holomorphic cross ratios
\begin{align}
0=&\frac{m_4  \epsilon _4 \sqrt{z\bar{z}}
   }{\Delta _4-1}\left(\left(\frac{\Delta _4-1}{2}+ (h_2 - h_1)
   z+ h_3\right) f^{h_1,h_2,h_3,\frac{\Delta
   _4-1}{2} }_{\bar{h}_1,\bar{h}_2,\bar{h}_3,\frac{\Delta
   _4-1}{2} }+
	(z-1) z\partial_z f^{h_1,h_2,h_3,\frac{\Delta_4-1}{2} }_{\bar{h}_1,\bar{h}_2,\bar{h}_3,\frac{\Delta_4-1}{2} }\right)\notag\\
	&+2 \epsilon _3  \sqrt{z\bar{z}}
   f^{h_1,h_2,h_3+\frac{1}{2},\frac{\Delta
   _4}{2}}_{\bar{h}_1,\bar{h}_2,\bar{h}_3+\frac{1}{2},\frac{\Delta
   _4}{2}}+2 z \epsilon _2
   f^{h_1,h_2+\frac{1}{2},h_3,\frac{\Delta
   _4}{2}}_{\bar{h}_1,\bar{h}_2+\frac{1}{2},\bar{h}_3,\frac{\Delta
   _4}{2}},\\
0=&\frac{m_4  \epsilon _4 \sqrt{z\bar{z}}
   }{\Delta _4-1}\left(\left(\frac{\Delta _4-1}{2}+ (\bar h_2 - \bar h_1)
   \bar z+ \bar h_3\right) f^{h_1,h_2,h_3,\frac{\Delta
   _4-1}{2} }_{\bar{h}_1,\bar{h}_2,\bar{h}_3,\frac{\Delta
   _4-1}{2} }+
	(\bar z-1) \bar z\partial_{\bar z} f^{h_1,h_2,h_3,\frac{\Delta_4-1}{2} }_{\bar{h}_1,\bar{h}_2,\bar{h}_3,\frac{\Delta_4-1}{2} }\right)\notag\\
	&+2 \epsilon _3 \sqrt{z \bar z}  
   f^{h_1,h_2,h_3+\frac{1}{2},\frac{\Delta
   _4}{2}}_{\bar{h}_1,\bar{h}_2,\bar{h}_3+\frac{1}{2},\frac{\Delta
   _4}{2}}+2 \bar z \epsilon _2
   f^{h_1,h_2+\frac{1}{2},h_3,\frac{\Delta
   _4}{2}}_{\bar{h}_1,\bar{h}_2+\frac{1}{2},\bar{h}_3,\frac{\Delta
   _4}{2}},
\end{align}
\begin{align}
0=&\frac{m_4 \epsilon _4 \sqrt{z \bar{z}}}{\Delta _4-1}\left(
\left(\frac{1-\Delta _4}{2} +(h_1 -h_2) z-h_3\right)
   \left(\frac{\Delta _4-1}{2}+\left(\bar{h}_2-\bar{h}_1\right)
   \bar{z}+\bar{h}_3 \right)
   f^{h_1,h_2,h_3,\frac{\Delta_4-1}{2} }_{\bar{h}_1,\bar{h}_2,\bar{h}_3,\frac{\Delta_4-1}{2} }
\right.\notag\\
&\left.+z \bar{z}(z-1) \left(\bar{z}-1\right)  \partial_z\partial_{\bar z}f^{h_1,h_2,h_3,\frac{\Delta _4-1}{2}
   }_{\bar{h}_1,\bar{h}_2,\bar{h}_3,\frac{\Delta _4-1}{2}
   }\right)+ m_4  \epsilon _4 \Delta _4 \sqrt{z\bar{z}}
   f^{h_1,h_2,h_3,\frac{\Delta_4+1}{2}}_{\bar{h}_1,\bar{h}_2,\bar{h}_3,\frac{\Delta_4+1}{2} }\notag\\
	&+\epsilon_2\left(2 \left(\Delta _1-\Delta _2+\Delta _4-1\right) z \bar{z}-z \left(2 \bar{h}_3+\Delta _4-1\right)-\bar{z} \left(\Delta _4+2 h_3-1\right)\right)f^{h_1,h_2+\frac{1}{2},h_3,\frac{\Delta
   _4}{2}}_{\bar{h}_1,\bar{h}_2+\frac{1}{2},\bar{h}_3,\frac{\Delta
   _4}{2}}\notag\\
	&+
	2  \epsilon _3 \sqrt{z\bar{z}}
   \left((\bar{h}_1-\bar{h}_2) \bar{z}+(h_1-h_2)
   z-\Delta_3\right) f^{h_1,h_2,h_3+\frac{1}{2},\frac{\Delta_4}{2}}_{\bar{h}_1,\bar{h}_2,\bar{h}_3+\frac{1}{2},\frac{\Delta_4}{2}},
\end{align}
where for brevity we suppressed the cross ratio dependence in function $f=f\left(z,\bar{z}\right)$.
Having a mix of recurrence and differential equations in two dimensions makes these constraints hard to solve. However, they may be a valuable aid to cross-check results when performing the map to the celestial sphere on explicit examples.

Since only one massive particle is participating in the scattering process, we can arrange momentum conservation to create an additional constraint
\begin{align}
(\epsilon_1 P_1+\epsilon_2 P_2+\epsilon_3 P_3)^{2}A_4^{0,0,0,m}=P_4^2 A_4^{0,0,0,m}
\end{align}
which leads to
\begin{align}
0=&\sqrt{z \bar{z}} \left(m^2 f_{\bar{h}_1,\bar{h}_2,\bar{h}_3,\frac{\Delta
   _4}{2}}^{h_1,h_2,h_3,\frac{\Delta
   _4}{2}}\left(z,\bar{z}\right)-4 \epsilon _1 \epsilon _2
   f_{\bar{h}_1+\frac{1}{2},\bar{h}_2+\frac{1}{2},\bar{h}_3,\frac{\Delta
   _4}{2}}^{h_1+\frac{1}{2},h_2+\frac{1}{2},h_3,\frac{\Delta
   _4}{2}}\left(z,\bar{z}\right)\right)\\
	&-4 \epsilon _3 \left((z-1)  \left(\bar{z}-1\right)\epsilon _2
   f_{\bar{h}_1,\bar{h}_2+\frac{1}{2},\bar{h}_3+\frac{1}{2},\frac{\Delta
   _4}{2}}^{h_1,h_2+\frac{1}{2},h_3+\frac{1}{2},\frac{\Delta
   _4}{2}}\left(z,\bar{z}\right)+\epsilon _1
   f_{\bar{h}_1+\frac{1}{2},\bar{h}_2,\bar{h}_3+\frac{1}{2},\frac{\Delta
   _4}{2}}^{h_1+\frac{1}{2},h_2,h_3+\frac{1}{2},\frac{\Delta
   _4}{2}}\left(z,\bar{z}\right)\right).\notag
\end{align}
 This constraint could be particularly useful, since it only involves shifts in conformal weights of the massless particles and no $z,\bar z$ derivatives.

\paragraph{Two massless, two massive and beyond:} In four-point cases when two or more of the scattering particles are massive scalars, the equations $\sum_{j} \epsilon_j P_j^{\mu}A_4=0$ always lead to four constraints involving shifts in conformal weights as well as up to second order derivatives in $z$ and $\bar z$; one constraint for each component $\mu$. This is conceptually different from the two- and three-point cases in which all coefficients of resulting monomials in $w_j,\bar w_j$ had to vanish separately, and it was not clear apriori how many constraints will arise. Since the four constraints can be generated straightforwardly from the explicit action of momentum generators $\sum_{j} \epsilon_j P_j^{\mu}A_4=0$ and are rather unwieldy, we omit them here.

\section{Discussion}
\label{sec:discussion}
An advantage of working with celestial amplitudes as opposed to amplitudes in Minkowski space, is that parameter restrictions such as (\ref{spin12eq}) automatically emerge as necessary conditions while solving Poincar\'e invariance constraints. To arrive at the same conclusions in Minkowski space, it is usually required to, e.g., additionally set up a basis of tensor structures and indirectly discover that certain scattering processes are ruled out since no corresponding tensor structure can be written for them (see, e.g., \cite{Arkani-Hamed:2017jhn}).

Considering that irreducible representations of the conformal group are labeled by conformal dimension $\Delta$ and spin $J$, different values of $\Delta$ and $J$ represent different operators in a CFT. On the other hand, irreducible representations of the Poincar\'e group are labeled by mass $m$ and spin $J$, so that the label $\Delta$ in this case can be varied for the same particle and therefore should be understood as a coordinate, or an energy reading, instead of a quantum number. This suggests that a Poincar\'e invariant theory on the celestial sphere should be noticeably different from a CFT. At the level of correlator structures we can observe the biggest differences for massless two-, three- and four-point structures where the structures become distribution-valued, while massive structures more readily resemble CFT correlators with a few extra conditions on the constants and parameters involved.

Additionally, as in \cite{Stieberger:2018onx}, we note that shifts in conformal dimensions $\Delta$, such as induced by the Poincar\'e momentum operator, take the dimension value off the complex line of continuous series representation $\Delta=1+i\lambda$ with $\lambda\in\mathbb{R}$. This breaks some of the CFT specific formalism that is employed in the literature. For instance, the formalism of conformal partial wave decomposition for four-point correlators \cite{Lam:2017ofc,Nandan:2019jas} relies on the partial wave orthogonality and convergence of the inner product employed, as well as the hermiticity of conformal Casimirs with respect to that inner product -- moving the conformal dimensions off the continuous series line generically breaks some of these properties. Related to our comment on the difference in irreducible representations of conformal versus Poincar\'e group, for partial wave considerations conformal Casimirs should in principle be replaced by Poincar\'e Casimirs; which suggests that in future work it may be interesting to instead derive a relativistic partial wave decomposition on the celestial sphere (see, e.g., \cite{Joos:1962qq,Macfarlane:1962zza}).

With these remarks we aim to emphasize that the applicability of CFT specific techniques in a Poincar\'e invariant theory has limitations.

\acknowledgments

MZ thanks D. Nandan, A. Schreiber, A. Volovich and C. Wen for discussions and collaboration on related topics. We are also grateful to M. Pate, A. Raclariu and A. Joyce for interesting discussions. AL and MZ are supported by the US Department of Energy under contract DE-SC0011941.

\appendix

\section{Three-point gluon celestial amplitude in Minkowski signature}
\label{sec:gluon3pt}
The three-point celestial amplitude for gluons was considered in \cite{Pasterski:2017ylz}, where the result was worked out in $(2,2)$ spacetime signature. This signature was chosen since the amplitude vanishes for physical (real) external momentum configurations in Minkowski signature $(1,3)$.

Here we revisit this amplitude in Minkowski signature and find that the result matches the distribution valued structure found in (\ref{struct3pt}). Naturally, the result still vanishes on the support of delta functions; however, being a parametrized zero, it allows us to extract the three point function coefficient for gluons.

As in \cite{Pasterski:2017ylz}, we consider the $\mathcal{A}^{--+}_3=\langle 12\rangle^3/(\langle 23\rangle\langle 31\rangle)$ three-point gluon amplitude with particles $1,2$ incoming and $3$ outgoing; where spinor helicity brackets are resolved as
\begin{align}
[ij]=2\sqrt{\omega_i\omega_j}\bar w_{ij}~~~,~~~\langle ij\rangle=-2\epsilon_i\epsilon_j\sqrt{\omega_i\omega_j} w_{ij},
\end{align}
with $w_{ij}=w_i-w_j,~\bar w_{ij}=\bar w_i-\bar w_j$. The map to the celestial sphere reads
\begin{align}
A_3^{--+}=\int_0^{\infty} \prod_{i=1}^3d\omega_i\omega_i^{\Delta_i-1}\mathcal{A}_3^{--+}\delta^{(4)}(\omega_1q^\mu_1+\omega_2q^\mu_2-\omega_3q^\mu_3).
\end{align}
The momentum conservation delta functions can be rewritten as follows
\begin{align}
\delta^{(4)}\left(\sum_{i=1}^3\epsilon_i\omega_iq^\mu_i\right)=\frac{1}{2\omega_1\omega_2(\omega_1+\omega_2)}&\delta(\omega_3-\omega_1-\omega_2)\delta\left(\frac{w_3+\bar w_3}{2}-\frac{w_1+\bar w_1}{2}\right)\delta\left(i\frac{\bar w_3-w_3}{2}-i\frac{\bar w_1-w_1}{2}\right)\notag\\
&\delta\left(\left(\frac{w_1+\bar w_1}{2}-\frac{w_2+\bar w_2}{2}\right)^2+\left(i\frac{\bar w_1-w_1}{2}-i\frac{\bar w_2-w_2}{2}\right)^2\right).
\end{align}
Note that the last delta function is equivalent to two delta functions separately setting the real and imaginary parts of $w_1-w_2$ to zero. This follows from the identity $\delta(x^2+y^2)=\frac{\pi}{2}\delta(x)\delta(y)$, which can be derived as\footnote{Where we use the half-range result $\int_0^\infty dx f(x)\delta(x)=\frac{1}{2}f(0)$.}
\begin{align}
\int dx dy f(x,y)\delta(x^2+y^2)=&\int d\theta dR~ R f(\sin\theta R,\cos\theta R)\delta(R^2)=\frac{\pi}{2} f(0,0)\notag\\
=&\frac{\pi}{2} \int dx dy f(x,y) \delta(x)\delta(y).
\end{align}
Rearranging linear combinations of delta function arguments for convenience leads to
\begin{align}
\delta^{(4)}\left(\sum_{i=1}^3\epsilon_i\omega_iq^\mu_i\right)=\frac{\pi \delta(\omega_3-\omega_1-\omega_2)}{2\omega_1\omega_2(\omega_1+\omega_2)}\delta\left(w_1-w_2\right)\delta\left(\bar w_1-\bar w_2\right)\delta\left(w_1-w_3\right)\delta\left(\bar w_1-\bar w_3\right),
\end{align}
where the delta functions are now in line with (\ref{struct3pt}).

The $\omega_3$ integration sets $\omega_3=\omega_1+\omega_2$, while $\omega_1,\omega_2$ integrations proceed as
\begin{align}
\int_0^\infty d\omega_1 d\omega_2 ~\omega_1^{i\lambda_1}\omega_2^{i\lambda_2}(\omega_1+\omega_2)^{i\lambda_3-2}=&\mathcal{B}(1+i\lambda_2,1-i\lambda_2-i\lambda_3)\int_0^\infty d\omega_1 \omega_1^{i\sum_j\lambda_j-1}\notag\\
=&2\pi\mathcal{B}(h_1+\bar h_1,h_2+\bar h_2)\delta(\sum_j\lambda_j),
\end{align}
where $\mathcal{B}(x,y)$ is the Euler beta function.

Overall, the celestial amplitude result amounts to
\begin{align}
A_3^{--+}=\pi^2\mathcal{B}(h_1+\bar h_1,h_2+\bar h_2)\delta(\sum_j\lambda_j)w_{12}^3w_{23}^{-1}w_{31}^{-1}\delta\left(w_{12}\right)\delta\left(\bar w_{12}\right)\delta\left(w_{13}\right)\delta\left(\bar w_{13}\right).
\end{align}
Making use of the equivalent transformations (\ref{trafo12}) on the support of the delta functions, we observe a match between the current result and (\ref{A3phase}) if we take $\delta(\sum_j\lambda_j)$ into account
\begin{align}
w_{12}^{3}w_{23}^{-1}w_{31}^{-1}~\to~ w_{12}=|w_{12}|\, e^{i\alpha}~\leftarrow~ \left| w_{12}\right| ^{1-i (\lambda _1+ \lambda _2+ \lambda _3)} e^{-i \alpha  \left(J_1+J_2+J_3\right)},
\end{align}
with $J_1=J_2=-1$ and $J_3=+1$.
This implies that the three-point gluon celestial amplitude vanishes for $w_{12}= 0$, as expected. However, with this we are still able to identify the corresponding three-point function coefficient for gluons
\begin{align}
\label{gluon3ptC}
C_{\bar h_1,\bar h_2,\bar h_3}^{h_1,h_2,h_3}=\pi^2\mathcal{B}(h_1+\bar h_1,h_2+\bar h_2)\delta(\sum_j\lambda_j).
\end{align}
Poincar\'e invariance demands that the three-point function coefficient satisfies the constraint (\ref{Pzero3ptC}), where here we have $\epsilon_1=\epsilon_2=-\epsilon_3$.
In this case, the constraint boils down to the relation $\mathcal{B}(x,y)=\mathcal{B}(x+1,y)+\mathcal{B}(x,y+1)$, which indeed is an identity exactly satisfied by the Euler beta function.

The calculation for the helicity flipped case $\mathcal{A}_3^{++-}$ is analogous, and leads to exactly the same three-point function coefficient, which always involves an Euler beta function dependent on $h_i,\bar h_i$ of the two incoming particles.

\section{Solving shift relations for massless four-point structure}
\label{A4shifts}
In this appendix we determine the generic class of functions satisfying the shift relations (\ref{A4fconstr2}) concerning the massless four-point structure.

Substituting the equations (\ref{A4fconstr2}) into each other in all possible ways, most generally we are considering the set of constraints
\begin{align}
\label{A4fconstr3}
e^{\partial_{\Delta_i}}\tilde f_{\Delta_1,\Delta_2,\Delta_3,\Delta_4}^{J_1,J_2,J_3,J_4}(z ,\bar z)=e^{\partial_{\Delta_j}}\tilde f_{\Delta_1,\Delta_2,\Delta_3,\Delta_4}^{J_1,J_2,J_3,J_4}(z ,\bar z),
\end{align}
for all $i,j\in\{1,2,3,4\}$ such that $i\neq j$. The constraints (\ref{A4fconstr3}) can be understood as the $n=4$ case of the following more general problem. 

Consider a function in $n$ variables $f(z_1,z_2,...,z_n)$ and impose the constraints
\begin{align}
\label{fzrel}
e^{\partial_{z_i}}f(z_1,z_2,...,z_n)=e^{\partial_{z_j}}f(z_1,z_2,...,z_n)~~~\text{for all pairs}~~~i\neq j\,,~i,j\in\{1,...,n\}.
\end{align}
We want to determine the general class of functions satisfying these constraints.
First, let us consider the first constraint
\begin{align}
f(z_1+1,z_2,...)=f(z_1,z_2+1,...).
\end{align}
Note that any function $f(x,y,...)$ can be equivalently expressed in terms of an appropriate new function $\tilde f(x+y,y,...)$, since both involved arguments stay linearly independent. In particular, $\tilde f(x+y,y,...)$ is obtained from $f(x,y,...)$ by writing $f(x+y-y,y,...)$ and absorbing the $-y$ from the first argument into the $y$ dependence of the second argument. 

Therefore, we replace $f(z_1,z_2,...,z_n)$ by an equivalent function $\tilde f(z_1+z_2,z_2,...,z_n)$ and relabel $\tilde f$ back to $f$ to avoid clutter of notation. We then find that the first constraint
\begin{align}
f(z_1+z_2+1,z_2,...)=f(z_1+z_2+1,z_2+1,...)
\end{align}
enforces periodicity of period $1$ for all functional dependence on $z_2$ which is not of the combined form $z_1+z_2$.

Now we continue and repeat the above step for equality of shifts $z_1\to z_1+1$ and $z_3\to z_3+1$, which in the same fashion leads to a new function $f(z_1+z_2+z_3,z_2,z_3,...,z_n)$ such that all functional dependence on $z_2$ or $z_3$, if not in combination $z_1+z_2+z_3$, must be periodic of period $1$.

Continuing this procedure $n-1$ times leads to $f(\sum_{i=1}^nz_i,z_2,z_3,...,z_n)$, such that all functional dependence on $z_2$ through $z_n$, which does not appear in combination $\sum_{i=1}^nz_i$, must be periodic of period $1$.

Finally, note that singling out the first variable $z_1$ to start the procedure was an arbitrary but fixed choice. We can repeat the above steps singling out, e.g., $z_2$ instead, and will find that any functional dependence on $z_1$, which does not appear in combination $\sum_{i=1}^nz_i$, must also be periodic with period $1$.

Therefore, we conclude that the general solution to the shift constraints (\ref{fzrel}) is given by a function $f(z_1,z_2,z_3,...,z_n)$ such that any functional $z_i$ dependence, which does not involve the combination $\sum_{i=1}^nz_i$, must be periodic of period $1$.

Thus, analogously, constraints (\ref{A4fconstr2}) and (\ref{A4fconstr3}) in general are solved by functions
\begin{align}
\tilde f_{\Delta_1,\Delta_2,\Delta_3,\Delta_4}^{J_1,J_2,J_3,J_4}(z ,\bar z),
\end{align}
where the functional $\Delta_i$ dependence must either appear as the combination $\sum_{i=1}^4\Delta_i$, or else be periodic of period $1$.

%


%

\end{document}